\documentclass[a4paper,12pt]{article}
\usepackage{makeidx}
\usepackage{amssymb}
\usepackage{graphicx}
\usepackage{amsmath}

\input{tcilatex}

\begin{document}

\begin{center}
{\LARGE Difference in Differences\ and Ratio in Ratios}

{\LARGE for Limited Dependent Variables}

(January 2023)\medskip

\begin{tabular}{l}
Sanghyeok Lee \\ 
Department of Economics \\ 
American University in Cairo \\ 
New Cairo 11835, Egypt \\ 
sanghyeok.lee@aucegypt.edu%
\end{tabular}%
\ \ \ \ \ \ \ 
\begin{tabular}{l}
Myoung-jae Lee (corresponding author) \\ 
Department of Economics, Korea University \\ 
145 Anam-ro, Sungbuk-gu \\ 
Seoul 02841, South Korea \\ 
myoungjae@korea.ac.kr%
\end{tabular}%
\bigskip \bigskip \bigskip \bigskip \bigskip \bigskip \bigskip
\end{center}

Difference in differences (DD) is widely used to find policy/treatment
effects with observational data, but applying DD to limited dependent
variables (LDV's)\textbf{\ }$Y$ has been problematic. This paper addresses
how to apply DD and related approaches (such as \textquotedblleft ratio in
ratios\textquotedblright\ or \textquotedblleft ratio in odds
ratios\textquotedblright ) to binary, count, fractional, multinomial or
zero-censored $Y$ under the unifying framework of `generalized linear models
with link functions'. We evaluate DD and the related approaches with
simulation and empirical studies, and recommend `Poisson Quasi-MLE' for
non-negative (such as count or zero-censored) $Y$ and (multinomial) logit
MLE for binary, fractional or multinomial $Y$.\bigskip \bigskip \bigskip
\bigskip

\textbf{Running Head}: DD and RR for LDV.\medskip

\medskip

\textbf{Data and Program Availability}: The data and programs used in this
paper are available from the first author upon request.\medskip

\textbf{Compliance with Ethical Standards \& No Conflict of Interest}:\ No
human or animal subject is involved in this research, and there is no
conflict of interest to disclose.\pagebreak

\section{Introduction}

\qquad Difference in Differences (DD) is one of the most popular research
designs in social sciences. Not just in social sciences, DD has been gaining
popularity also in natural sciences, as can be seen in, e.g., Jena et al.
(2015), Cataife and Pagano (2017), and McGrath et al. (2019). There are
various references for DD: Angrist and Krueger (1999), Shadish et al.
(2002), Lee (2005, 2016a), Athey and Imbens (2006), Angrist and Pischke
(2009), Lechner (2011), Lee and Kim (2014), Morgan and Winship (2014), Kim
and Lee (2017), Lee and Sawada (2020), Kahn-Lang and Lang (2020), etc.

\qquad DD is basically for linear models with additive components, which
makes applying DD to limited dependent variables (LDV's) with nonlinear
models problematic. This paper provides answers to this problem, using the
unifying idea of `generalized linear models with link functions'.

\qquad Consider an outcome/response $Y_{it}$ for subject $i$ at time $t=2,3$%
, a time-constant treatment qualification dummy $Q_{i}$, and a binary
treatment $D_{it}$; we set $t=2,3$ to avoid the confusion with dummy
variable values $0,1$. The hallmark of DD is that $D_{it}$ is the
interaction of $Q_{i}$ and $1[t=3]$: $D_{it}=Q_{i}1[t=3]$, where $1[A]\equiv
1$ if $A$ holds and $0$ otherwise. That is, only the $Q_{i}=1$ group is
treated at $t=3$, and untreated otherwise.

\qquad DD can be implemented with panel data or repeated cross-sections
(RCS). We use RCS in this paper, because RCS are easier to collect than
panel data and also because our empirical study uses RCS\textbf{.} In
typical RCS, there is a huge reservoir of subjects, and random sampling for
a sample size $N$ is done each period. Hence, we can safely assume that each
subject is sampled only once in RCS, and that the sampling dummy $S_{i}$\ is
independent of the other random variables;%
\begin{equation*}
D_{i}=Q_{i}S_{i}\ \ \ \ \ \text{where}\ \ \ \ \ S_{i}\equiv 1[i\text{ is
sampled at }t=3].
\end{equation*}%
Let $Y_{it}^{d}$ be the potential version of $Y_{it}$ for $D_{it}=d=0,1$,
and $Y_{i}^{d}=(1-S_{i})Y_{i2}^{d}+S_{i}Y_{i3}^{d}\ $be the RCS potential
response. Let $W_{it}$ denote covariates\textbf{, }and $W_{i}\equiv
(1-S_{i})W_{i2}+S_{i}W_{i3}$ be the RCS covariates. Clearly, RCS variables
are derived from the underlying panel model variables. Henceforth, we often
omit the subscript $i$ indexing subjects.

\qquad As a preliminary, ignoring the covariates $W$ for a while, define for
RCS:%
\begin{equation}
\mu _{QS}\equiv E(Y|Q,S)=\lambda ^{-1}(\beta _{2}+\beta _{\tau }S+\beta
_{q}Q+\beta _{d}D),\ \ \ \ \text{\ }\beta _{\tau }\equiv \beta _{3}-\beta
_{2},  \tag{1.1}
\end{equation}%
where $\lambda (\cdot )$ is a `link function' as\ in the generalized linear
model (Nelder and Wedderburn 1972), $(\beta _{2},\beta _{3})$ are the period-%
$(2,3)$ intercepts, $\beta _{\tau }$ is the time effect of $t=3$ relative to 
$t=2$, $\beta _{q}$ is the group effect of $Q=1$, and $\beta _{d}$ is the
desired treatment effect.

\qquad Since $(Q,S)$ generates four cells for the four parameters $(\beta
_{2},\beta _{\tau },\beta _{q},\beta _{d})$ in (1.1), there seems no loss of
generality in (1.1). However, (1.1) does include a restriction: $QS$ should
not appear separately from the treatment $D$. If the group effect of $Q$
changes across time, then $QS$ becomes relevant other than through $D$. This
restriction---no change in the group effect over time---is the well-known DD
`parallel trend assumption'.\bigskip

\qquad For continuous $Y$, $\lambda (\cdot )$ in (1.1) is the identity, so
that $\mu _{QS}=\beta _{2}+\beta _{\tau }S+\beta _{q}Q+\beta _{d}D$. For
this, DD is%
\begin{equation*}
\mu _{11}-\mu _{10}-(\mu _{01}-\mu _{00})=(\beta _{\tau }+\beta _{d})-\beta
_{\tau }=\beta _{d}:
\end{equation*}%
DD removes $\beta _{2}+\beta _{\tau }S+\beta _{q}Q$ to leave $\beta _{d}D$
that changes across both times and groups. In practice, to account for the
covariates $W$, a linear model such as%
\begin{equation}
E(Y|Q,S,W)=\beta _{2}+\beta _{\tau }S+\beta _{q}Q+\beta _{d}D+\beta
_{w}^{\prime }W  \tag{1.2}
\end{equation}%
is estimated to find the slope of $D$ as the treatment effect.

\qquad For LDV's, the story changes much. E.g., consider $Y=1[0\leq Y^{\ast
}]$ where $Y^{\ast }$ is the latent continuous outcome. With the $N(0,1)$
distribution function $\Phi (\cdot )$, the probit is%
\begin{equation}
E(Y|Q,S)=P(Y=1|Q,S)=\Phi (\beta _{2}+\beta _{\tau }S+\beta _{q}Q+\beta
_{d}D).  \tag{1.3}
\end{equation}%
One way to stick to DD is estimating (1.3) to interpret $\beta _{d}$ as the
effect on $Y^{\ast }$, not on $Y$. E.g., if $\beta _{d}=2$, then $D$ shifts $%
Y^{\ast }$ by twice the standard deviation (SD) of $Y^{\ast }$. However,
many practitioners desire the effect as a change in $P(Y=1|Q,S)$, not in $%
Y^{\ast }$.

\qquad The `marginal effect' that is a change of $P(Y=1|Q,S)$ in (1.3)\ due
to $D$ is%
\begin{equation}
\Phi (\beta _{2}+\beta _{\tau }S+\beta _{q}Q+\beta _{d})-\Phi (\beta
_{2}+\beta _{\tau }S+\beta _{q}Q).  \tag{1.4}
\end{equation}%
Ai and Norton (2003) noted that this is not the correct effect, but their
criticism applies to the case of an interaction treatment, where both $Q$
and $S$ are genuine treatments and the interest is in the effect of taking
both treatments (e.g., drugs) together. Differently from this, $Q$ and $S$
are not treatments per se in the usual DD, and $D=QS$ just happened to be
the way the treatment was implemented. Indeed, Puhani (2012, eq. (10))
showed that (1.4) with $S=Q=1$ is a legitimate treatment effect of interest.

\qquad The complication involving (1.3) and (1.4) arises because DD is
applied to a nonlinear model, despite that DD is appropriate for linear
models. To drive home our point, consider the `log link' $\lambda (\cdot
)=\ln (\cdot )\Longleftrightarrow \lambda ^{-1}(\cdot )=\exp (\cdot )$, with
which (1.1) becomes%
\begin{equation}
\mu _{QS}\equiv E(Y|Q,S)=\exp (\beta _{2}+\beta _{\tau }S+\beta _{q}Q+\beta
_{d}D).  \tag{1.5}
\end{equation}%
This is appropriate for non-negative responses. For (1.5), `ratio in ratios
(RR)' removes the time and group effects, and `RR minus one' gives the
proportional effect of $D$:%
\begin{equation}
\frac{\mu _{11}}{\mu _{10}}/(\frac{\mu _{01}}{\mu _{00}})-1=\exp (\beta
_{d})-1.  \tag{1.6}
\end{equation}%
In practice, as the linear model (1.2) is used instead of the DD $\mu
_{11}-\mu _{10}-(\mu _{01}-\mu _{00})$ to find $\beta _{d}$, $\exp (\beta
_{2}+\beta _{\tau }S+\beta _{q}Q+\beta _{d}D+\beta _{w}^{\prime }W)$ is used
instead of the RR in (1.6).\bigskip

\qquad This paper makes the following contributions, some of which might
have been known, although we cannot point out the exact references as DD has
been applied widely. First, we adopt the unifying framework of generalized
linear models with link functions. Second, we advocate RR for non-negative
(such as count or zero-censored) responses based on the log link, and
\textquotedblleft ratio in odds ratios (ROR)\textquotedblright\ for binary,
fractional or multinomial responses based on the `logit link'. Third,
although ROR is difficult to interpret, we show that it becomes a
proportional effect for \textquotedblleft rare events\textquotedblright .
Fourth, if more than two periods are available, we propose a simple test for
the DD parallel trends and analogous assumptions for RR and ROR: test for
zero slope of $tQ$, as\ zero slope validates the DD parallel trends and
analogous assumptions for RR and ROR.

\qquad Practitioners often ignore the LDV\ nature of $Y$, and simply use a
linear model for DD. One justification for this was provided by Lee (2018)
for any response $Y$ and a binary exogenous $D$: under $(Y^{0},Y^{1})\amalg
D|W$ with `$\amalg $' for independence, it holds that%
\begin{equation}
Y=E(Y^{0}|W)+E(Y^{1}-Y^{0}|W)D+error\text{.}  \tag{1.7}
\end{equation}%
This representation for exogenous $D$ was generalized for endogenous $D$ in
Lee (2021). Then, justifying a linear model for LDV's can be done by
linearly approximating $E(Y^{0}|W)$ and $E(Y^{1}-Y^{0}|W)$ in (1.7)\textbf{.}
However, this paper's approach is using LDV's as such without such
approximations.

\qquad In the remainder of this paper, Sections 2 and 3 examine RR and ROR,
respectively, where the covariates are controlled in addition to $(Q,S)$.
Section 4 presents an empirical analysis for various health outcomes.
Section 5 concludes this paper. The appendix contains proofs, as well as a
simulation study to show that the usual linear-model DD is misleading for
LDV's whereas RR and ROR approaches work well.

\section{Ratio in Ratios (RR) for Non-Negative Response}

\qquad This section studies RR for non-negative responses including count
and zero-censored responses. First, the identification aspect is examined.
Second, although RR can be estimated nonparametrically replacing the
conditional means in RR with sample analogs, this is not how RR would be
estimated in practice; instead, a practical semiparametric estimator for RR
is advocated. Third, several remarks are made.

\subsection{Proportional Effect Identification with RR}

\qquad To simplify notation when covariates $W$ are allowed for, define%
\begin{equation}
\mu _{QS}(w)\equiv E(Y|w,Q,S)  \tag{2.1}
\end{equation}

where $E(Y|w,Q,S)$ is a shorthand for $E(Y|W=w,Q,S)$. With this, define RR\
conditional on $W=w$ analogously to (1.6) plus one:%
\begin{equation*}
RR(w)\equiv \frac{\mu _{11}(w)}{\mu _{10}(w)}/\{\frac{\mu _{01}(w)}{\mu
_{00}(w)}\}.
\end{equation*}

\qquad The identification condition for $RR(w)$ is%
\begin{equation}
\left( \frac{E(Y_{3}^{0}|w,Q=1)}{E(Y_{2}^{0}|w,Q=1)}\right) /\left( \frac{%
E(Y_{3}^{0}|w,Q=0)}{E(Y_{2}^{0}|w,Q=0)}\right) =1;  \tag{ID$_{RR}$}
\end{equation}%
keep in mind that $S$ is independent of the other random variables, and ID$%
_{RR}$ involves only untreated responses. In ID$_{RR}$, $E(Y_{3}^{0}|w,Q=1)$
is a counterfactual, because only $E(Y_{3}^{1}|w,Q=1)$ is realized for $Q=1$
at $t=3$. ID$_{RR}$ is analogous to the usual DD identification condition
(i.e., parallel trends) appropriate for linear models:%
\begin{equation}
E(Y_{3}^{0}|w,Q=1)-E(Y_{2}^{0}|w,Q=1)-%
\{E(Y_{3}^{0}|w,Q=0)-E(Y_{2}^{0}|w,Q=0)\}=0.  \tag{ID$_{DD}$}
\end{equation}

\qquad The main point is that $RR(w)-1$ is the `proportional effect on the
treated at the post-treatment period $t=3$', in view of the first and last
expressions of the following:%
\begin{eqnarray}
&&RR(w)-1=\left( \frac{E(Y|w,Q=1,S=1)}{E(Y|w,Q=1,S=0)}\right) /\left( \frac{%
E(Y|w,Q=0,S=1)}{E(Y|w,Q=0,S=0)}\right) \ -\ 1  \notag \\
&&\ =\left( \frac{E(Y_{3}^{1}|w,Q=1)}{E(Y_{2}^{0}|w,Q=1)}\right) /\left( 
\frac{E(Y_{3}^{0}|w,Q=0)}{E(Y_{2}^{0}|w,Q=0)}\right) \ -\ 1\ \ \ \text{(as }%
S\amalg \text{`the other variables')}  \notag \\
&&\ =\frac{E(Y_{3}^{1}|w,Q=1)}{E(Y_{3}^{0}|w,Q=1)}\cdot \left( \frac{%
E(Y_{3}^{0}|w,Q=1)}{E(Y_{2}^{0}|w,Q=1)}\right) /\left( \frac{%
E(Y_{3}^{0}|w,Q=0)}{E(Y_{2}^{0}|w,Q=0)}\right) \ -\ 1  \notag \\
&&\ =\frac{E(Y_{3}^{1}|w,Q=1)}{E(Y_{3}^{0}|w,Q=1)}-1=\frac{%
E(Y_{3}^{1}-Y_{3}^{0}|w,Q=1)}{E(Y_{3}^{0}|w,Q=1)}\text{ \ \ \ \ (under ID}%
_{RR}\text{).}  \TCItag{2.2}
\end{eqnarray}

\qquad If the dimension of $W$ is low (or if $W$ is discrete), $RR(w)$ can
be estimated nonparametrically by substituting nonparametric estimators into
the four components of $RR(w)$. In practice, however, typically the
dimension of $W$ is high, and thus we explore a simpler semiparametric
exponential regression next---semiparametric because only $E(Y|W,Q,S)$ is
specified, not the full distribution of $Y|(W,Q,S)$.

\subsection{Poisson Quasi-MLE (QMLE)}

\qquad In view of (1.5), suppose that a panel data exponential model holds
for $Y_{it}^{d}$:%
\begin{eqnarray}
&&E(Y_{t}^{d}|W_{t},Q)=\exp (\beta _{t}+\beta _{q}Q+\beta
_{d}d+W_{t}^{\prime }\beta _{w})  \TCItag{2.3} \\
&\Longleftrightarrow &\ Y_{t}^{d}=\exp (\beta _{t}+\beta _{q}Q+\beta
_{d}d+W_{t}^{\prime }\beta _{w}+U_{t}),\text{ \ \ }\exp (U_{t})\equiv
Y_{t}^{d}/E(Y_{t}^{d}|W_{t},Q);  \notag
\end{eqnarray}%
$\beta _{t}$ is a time-varying intercept, and $E\{\exp (U_{t})|W_{t},Q\}=1$
holds. ID$_{RR}$ holds for (2.3):%
\begin{eqnarray*}
&&\left( \frac{E(Y_{3}^{0}|w,Q=1)}{E(Y_{2}^{0}|w,Q=1)}\right) /\left( \frac{%
E(Y_{3}^{0}|w,Q=0)}{E(Y_{2}^{0}|w,Q=0)}\right) \\
&=&\left( \frac{\exp (\beta _{3}+\beta _{q}+w^{\prime }\beta _{w})}{\exp
(\beta _{2}+\beta _{q}+w^{\prime }\beta _{w})}\right) /\left( \frac{\exp
(\beta _{3}+w^{\prime }\beta _{w})}{\exp (\beta _{2}+w^{\prime }\beta _{w})}%
\right) =1.
\end{eqnarray*}

\qquad Turn the panel data model (2.3) into the RCS model for $Y^{d}\equiv
(1-S)Y_{2}^{d}+SY_{3}^{d}$:%
\begin{eqnarray}
&&Y^{d}=\exp (\beta _{2}+\beta _{\tau }S+\beta _{q}Q+\beta _{d}d+W^{\prime
}\beta _{w}+U),  \TCItag{2.4} \\
&&\beta _{\tau }\equiv \beta _{3}-\beta _{2},\ \ \ W\equiv
(1-S)W_{2}+SW_{3},\ \ \ U\equiv (1-S)U_{2}+SU_{3};  \notag
\end{eqnarray}%
$(W,Q,S)$ is exogenous to $U$ in the sense $E\{\exp (U)|W,Q,S\}=1$. Take $%
E(\cdot |W,Q,S)$ on the observed $Y=(1-D)Y^{0}+DY^{1}$: due to $D=QS$,%
\begin{eqnarray}
&&E(Y|W,Q,S)=(1-QS)\cdot E(Y^{0}|W,Q,S)+QS\cdot E(Y^{1}|W,Q,S)  \notag \\
&&\ =\exp (\beta _{2}+\beta _{\tau }S+\beta _{q}Q+\beta _{d}D+W^{\prime
}\beta _{w})  \TCItag{2.5} \\
&&\ \Longrightarrow \ RR(w)-1=\exp (\beta _{d})-1\text{;}  \TCItag{2.6}
\end{eqnarray}%
the second equality can be verified by substituting $D=QS=0,1$ into both
sides of the equality. We use RCS and the model (2.5) to estimate $\beta _{d}
$ and other parameters.

\qquad The constant treatment effect $\beta _{d}$ can be easily allowed to
be a function of $W$, as in $\beta _{d}(W_{t})=\beta _{d0}+\beta
_{dw}^{\prime }W_{t}$ for parameters $(\beta _{d0},\beta _{dw})$. Then we
have $\beta _{d}(W)=\beta _{d0}+\beta _{dw}^{\prime }W$ in RCS, and (2.5)
becomes%
\begin{eqnarray*}
&&E(Y|W,Q,S)=\exp \{\beta _{2}+\beta _{\tau }S+\beta _{q}Q+\beta
_{d}(W)D+W^{\prime }\beta _{w}\} \\
&&\ \Longrightarrow \ RR(w)-1=\exp \{\beta _{d}(w)\}-1\text{ \ \ \ \
(proportional effect at }W=w\text{).}
\end{eqnarray*}

\qquad For estimation, the simplest approach is the `Poisson Quasi-Maximum
Likelihood Estimator (Poisson QMLE)'. The Poisson QMLE is the same as the
Poisson MLE, except that the variance is estimated with a \textquotedblleft
sandwich-form\textquotedblright\ asymptotic variance estimator. The maximand
for the Poisson QMLE is the same as that for Poisson MLE:%
\begin{equation*}
\sum_{i}\{Y_{i}(X_{i}^{\prime }b)-\exp (X_{i}^{\prime }b)\},\ \ \ \ \
X_{i}\equiv (1,S_{i},Q_{i},D_{i},W_{i}^{\prime })^{\prime },\text{ \ \ }%
b=(b_{2},b_{\tau },b_{q},b_{d},b_{w}^{\prime })^{\prime }.
\end{equation*}

\qquad The first order-condition at $b=\beta $ is $\sum_{i}\{Y_{i}-\exp
(X_{i}^{\prime }\beta )\}X_{i}=0$ where $\beta \equiv (\beta _{2},\beta
_{\tau },\beta _{q},\beta _{d},\beta _{w}^{\prime })^{\prime }$, which holds
due to $E(Y|X)=\exp (X^{\prime }\beta )$. The maximum is unique as the
second order derivative $-\sum_{i}X_{i}X_{i}^{\prime }\exp (X_{i}^{\prime
}b) $ is n.d.: just under $E(Y|X)=\exp (X^{\prime }\beta )$, the Poisson
QMLE\ is consistent for $\beta $. The Poisson QMLE for exponential models
was advocated in Lee (2005) and Santos Silva and Tenreyro (2006). For
heterogeneous effects, we may use $\beta _{d}(W)=\beta _{d0}+\beta
_{dw}^{\prime }W$ in the Poisson QMLE.

\subsection{Remarks}

\qquad Here we make a few remarks on the applicability of the above RR\
identification and Poisson QMLE to count and zero-censored responses. Bear
in mind that the semiparametric exponential regression model (2.5) requires
no upper bound on $Y$.

\qquad \textbf{First}, instead of the difference effect $%
E(Y_{3}^{1}-Y_{3}^{0}|w,Q=1)$, examining the proportional effect in (2.2)
can be beneficial (Yadlowsky et al. 2021). E.g., suppose $%
E(Y_{3}^{0}|w,Q=1)=G(w)$ for a function $G(\cdot )$ and the proportional
effect is a constant $\beta _{d}$. Then the difference effect $%
E(Y_{3}^{1}-Y_{3}^{0}|w,Q=1)=\beta _{d}G(w)$ introduces effect heterogeneity
unnecessarily, compared with the simple $\beta _{d}$. Proportional effects
for exponential models have been advocated in many studies: Lee and
Kobayashi (2001), Dukes and Vansteelandt (2018) and Ciani and Fisher (2019),
among others.\bigskip

\qquad \textbf{Second}, suppose $Y=\exp (Y^{\ast })$, $Y^{\ast }\equiv \beta
_{2}+\beta _{\tau }S+\beta _{q}Q+\beta _{d}D+W^{\prime }\beta _{w}+U$ and $%
E\{\exp (U)|W,Q,S\}=1$. Then we can interpret $\beta _{d}$ as the DD effect
on $Y^{\ast }$, whereas $\exp (\beta _{d})-1$ is the proportional effect on $%
Y$. However, for count responses such as $Y|(W,Q,S)$ generated by the
Poisson distribution with $P(Y=y|X)\ =\{\exp (X^{\prime }\beta )\}^{y}\exp
\{-\exp (X^{\prime }\beta )\}/y!$, there is no $Y^{\ast }$. In this Poisson
case, the proportional effect interpretation on the observed $Y$\ with RR is
the only way to meaningfully interpret the slope $\beta _{d}$\ of $D$\ in
the exponential model. This statement applies also to count responses based
on other distributions such as Negative Binomial.\bigskip

\qquad \textbf{Third}, if $\beta _{q\tau }tQ$ with $\beta _{q\tau }\neq 0$
appears as a regressor, then ID$_{RR}$ fails due to $\beta _{q\tau }tQ$:%
\begin{equation}
\left( \frac{\exp (\beta _{3}+\beta _{q}+3\beta _{q\tau }+w^{\prime }\beta
_{w})}{\exp (\beta _{2}+\beta _{q}+2\beta _{q\tau }+w^{\prime }\beta _{w})}%
\right) /\left( \frac{\exp (\beta _{3}+w^{\prime }\beta _{w})}{\exp (\beta
_{2}+w^{\prime }\beta _{w})}\right) =\exp (\beta _{q\tau })\neq 1.  \tag{2.7}
\end{equation}%
Hence, using $tQ$ as an extra regressor is an easy way to test or allow for
non-parallel trends or analogous conditions for RR. However, $tQ$ cannot be
used if only two periods are available, because using $tQ$ is equivalent to
using $QS$ which is $D$. With more than two periods available, there are two
ways to entertain $\beta _{q\tau }tQ$ as follows.

\qquad The first way is using $tQ$ as an extra regressor. For panel data, $%
tQ $ can be used as such, but for RCS, $Q_{i}^{\tau }\equiv
Q_{i}\sum_{t}S_{it}t $ should be used instead, where $S_{it}=1$ if $i$ is
sampled in period $t$ and $0$ otherwise. Intuitively speaking, the untreated
group difference is allowed to change linearly with $tQ$ over time, and then
any deviation from the change is taken as the treatment effect. With more
periods, the allowed linear untreated trend difference can be expanded to$\ $%
quadratic ($t^{2}Q$), cubic ($t^{3}Q$), and so on.

\qquad\ The other way is using triple ratios, or \textquotedblleft ratio in
ratios in ratios (RRR)\textquotedblright\ generalizing RR, analogously to
triple differences (Lee 2016b) to allow for non-parallel trends in DD. With $%
t=1,2,3$ available, let $m_{Qt}(w)\equiv E(Y|w,Q,\ $sampled at $t)$ and%
\begin{equation}
\left( \frac{E(Y_{t}^{0}|w,Q=1)}{E(Y_{t-1}^{0}|w,Q=1)}\right) /\left( \frac{%
E(Y_{t}^{0}|w,Q=0)}{E(Y_{t-1}^{0}|w,Q=0)}\right) =\gamma \text{ \ \ \ \ for }%
t=2,3\text{,}  \tag{ID$_{RRR}$}
\end{equation}%
which allows ID$_{RR}$ to be violated when $\gamma \neq 1$ as follows.
Observe%
\begin{eqnarray*}
&&RRR(w)\equiv \left[ \frac{m_{13}(w)}{m_{12}(w)}/\left\{ \frac{m_{03}(w)}{%
m_{02}(w)}\right\} \right] /\left[ \frac{m_{12}(w)}{m_{11}(w)}/\left\{ \frac{%
m_{02}(w)}{m_{01}(w)}\right\} \right] \\
&&\ =\left[ \left( \frac{E(Y_{3}^{1}|w,Q=1)}{E(Y_{2}^{0}|w,Q=1)}\right)
/\left( \frac{E(Y_{3}^{0}|w,Q=0)}{E(Y_{2}^{0}|w,Q=0)}\right) \right] \\
&&\ \ \ \ \ \ \ /\left[ \left( \frac{E(Y_{2}^{0}|w,Q=1)}{E(Y_{1}^{0}|w,Q=1)}%
\right) /\left( \frac{E(Y_{2}^{0}|w,Q=0)}{E(Y_{1}^{0}|w,Q=0)}\right) \right]
\\
&&\ =\frac{E(Y_{3}^{1}|w,Q=1)}{E(Y_{3}^{0}|w,Q=1)}\cdot \left[ \left( \frac{%
E(Y_{3}^{0}|w,Q=1)}{E(Y_{2}^{0}|w,Q=1)}\right) /\left( \frac{%
E(Y_{3}^{0}|w,Q=0)}{E(Y_{2}^{0}|w,Q=0)}\right) \right] \\
&&\ \ \ \ \ \ \ \ \ \ \ \ \ \ \ \ \ \ \ \ \ \ \ \ \ \ \ \ \ /\left[ \left( 
\frac{E(Y_{2}^{0}|w,Q=1)}{E(Y_{1}^{0}|w,Q=1)}\right) /\left( \frac{%
E(Y_{2}^{0}|w,Q=0)}{E(Y_{1}^{0}|w,Q=0)}\right) \right] .
\end{eqnarray*}%
The last two terms in $[\cdot ]$ are both equal to $\gamma $ to cancel each
other. Hence, under ID$_{RRR}$, RRR identifies the same effect as RR
identifies, even when ID$_{RR}$ fails.\bigskip

\qquad \textbf{Fourth}, consider a RCS zero-censored model:%
\begin{eqnarray}
&&Y=\max (0,Y^{\ast })=Y^{\ast }1[0<Y^{\ast }],\text{ \ \ }Y^{\ast }\equiv
\beta _{2}+\beta _{\tau }S+\beta _{q}Q+\beta _{d}D+W^{\prime }\beta _{w}+U 
\notag \\
&&\ \Longrightarrow \ E(Y|X)=E(Y^{\ast }1[0<Y^{\ast }]|X).  \TCItag{2.8}
\end{eqnarray}%
Since $E(Y|X)=E(Y^{\ast }1[0<Y^{\ast }]|X)$ is non-negative without any
upper bound, the exponential regression model (2.5) can be adopted, although
it may not be as appealing as for count responses because the transformation 
$\max (0,\cdot )$ is not smooth.

\qquad Santos Silva and Tenreyro (2011) showed that the exponential
regression holds for (2.8) if $Y_{i}=\sum_{j=1}^{M_{i}}Z_{ij}$, where $M_{i}$
is a non-negative integer random variable such as Poisson count, and $%
(Z_{i1},Z_{i2},...)$ are independent and identically distributed (iid)
positive random variables with $Z_{ij}\amalg M_{i}|X_{i}$; $Y=0$ occurs if $%
M=0$. Due to $Z_{ij}\amalg M_{i}|X_{i}$,%
\begin{eqnarray}
&&E(Y|X)=E(M|X)E(Z_{j}|X)  \notag \\
&&\ =\exp \{\alpha _{2}+\beta _{2}+(\alpha _{\tau }+\beta _{\tau })S+(\alpha
_{q}+\beta _{q})Q+\beta _{d}D+W^{\prime }(\alpha _{w}+\beta _{w})\} 
\TCItag{2.9} \\
&&\ \ \ \text{if \ \ }E(M|X)=\exp (\alpha _{2}+\alpha _{\tau }S+\alpha
_{q}Q+W^{\prime }\alpha _{w}),  \notag \\
&&\ \ \ \ \ \text{ \ \ }E(Z_{j}|X)=\exp (\beta _{2}+\beta _{\tau }S+\beta
_{q}Q+\beta _{d}D+W^{\prime }\beta _{w})\text{.}  \notag
\end{eqnarray}%
It is not clear what $Y^{\ast }$ is here, but the interpretation of $\exp
(\beta _{d})-1$ as a proportional effect on $Y$ still holds regardless of
what $Y^{\ast }$ might be.

\qquad A DD example for $Y_{i}=\sum_{j=1}^{M_{i}}Z_{ij}$ is that $Y_{i}$ is
the expenditure on tobacco by person $i$ in a year, $Z_{ij}$ is the tobacco
expenditure of person $i$ on day $j$, $M_{i}$ is the number of the
tobacco-purchasing days for person $i$ in the year, $Q_{i}=1$ if person $i$
is legally eligible to smoke, $W_{i}$ is individual traits of person $i$,
and there is a smoking-discouraging policy $D_{i}=Q_{i}S_{i}$ implemented at 
$t=3$ effectively increasing tobacco product prices. In this case, $M_{i}$
is how frequently tobacco is purchased which is unlikely to be affected by
the policy, and $Z_{ij}$ is the day-$j$ purchase amount affected by the
policy.

\section{Ratio in Odds-Ratios (ROR)}

\qquad This section studies ROR: we examine the identification aspect first,
followed by logit-based estimation for binary and fractional responses. ROR
is also applicable to multinomial response, but it is presented (along with
a simulation study) in the appendix due to the complexity involving multiple
equations and additional notation.

\subsection{Proportional Odds Effect Identification with ROR}

\qquad For binary $Y$, define the `odds conditional on $(W=w,Q=q,S=s)$' for
RCS as%
\begin{eqnarray}
R_{qs}(Y;w) &\equiv &\frac{P(Y=1|w,Q=q,S=s)}{P(Y=0|w,Q=q,S=s)}\text{ \ \ \ \
which leads to}  \TCItag{3.1} \\
R_{11}(Y;w) &=&R_{11}(Y_{3}^{1};w),\text{ \ \ \ \ }%
R_{01}(Y;w)=R_{01}(Y_{3}^{0};w),  \notag \\
R_{10}(Y;w) &=&R_{10}(Y_{2}^{0};w),\ \ \ \ \ R_{00}(Y;w)=R_{00}(Y_{2}^{0};w).
\notag
\end{eqnarray}%
Also define `Ratio in Odds-Ratios (ROR) conditional on $W=w$':%
\begin{equation*}
ROR(Y;w)\equiv \left( \frac{R_{11}(Y;w)}{R_{10}(Y;w)}\right) /\left( \frac{%
R_{01}(Y;w)}{R_{00}(Y;w)}\right) .
\end{equation*}

\qquad The identification condition to be invoked for ROR is%
\begin{equation}
ROR(Y^{0};w)=\left( \frac{R_{11}(Y_{3}^{0};w)}{R_{10}(Y_{2}^{0};w)}\right)
/\left( \frac{R_{01}(Y_{3}^{0};w)}{R_{00}(Y_{2}^{0};w)}\right) =1, 
\tag{ID$_{ROR}$}
\end{equation}%
where $R_{11}(Y_{3}^{0};w)$ is a counterfactual, because only $%
R_{11}(Y_{3}^{1};w)$ is realized. Doing analogously to (2.2), $ROR(Y;w)-1$
is the `proportional odds effect on the treated at the post-treatment period 
$t=3$'---`on the treated' because $R_{11}$ is for $Q=1$ and $S=1$:%
\begin{eqnarray}
&&ROR(Y;w)-1=\left( \frac{R_{11}(Y;w)}{R_{10}(Y;w)}\right) /\left( \frac{%
R_{01}(Y;w)}{R_{00}(Y;w)}\right) \ -\ 1  \notag \\
&=&\left( \frac{R_{11}(Y_{3}^{1};w)}{R_{10}(Y_{2}^{0};w)}\right) /\left( 
\frac{R_{01}(Y_{3}^{0};w)}{R_{00}(Y_{2}^{0};w)}\right) \ -\ 1  \notag \\
&=&\frac{R_{11}(Y_{3}^{1};w)}{R_{11}(Y_{3}^{0};w)}\cdot \left( \frac{%
R_{11}(Y_{3}^{0};w)}{R_{10}(Y_{2}^{0};w)}\right) /\left( \frac{%
R_{01}(Y_{3}^{0};w)}{R_{00}(Y_{2}^{0};w)}\right) \ -\ 1  \notag \\
&=&\frac{R_{11}(Y_{3}^{1};w)}{R_{11}(Y_{3}^{0};w)}-1=\frac{%
R_{11}(Y_{3}^{1};w)-R_{11}(Y_{3}^{0};w)}{R_{11}(Y_{3}^{0};w)}\ \ \ \ \ \text{%
(under ID}_{ROR}\text{).}  \TCItag{3.2}
\end{eqnarray}

\qquad One disadvantage of ROR compared with RR is the difficulty in
interpreting the `proportional odds effect'. For this, suppose $Y=1$ is a 
\textit{rare event} in the sense%
\begin{equation}
\frac{P(Y_{3}^{0}=0|w,Q=1)}{P(Y_{3}^{1}=0|w,Q=1)}\simeq 1\text{ \ \ \ \ for
all }w\text{;}  \tag{3.3}
\end{equation}%
e.g., $Y=1$ is a rare cancer occurrence such that $P(Y_{3}^{d}=0|w,Q=1)%
\simeq 1$ for all $w$ and $d=0,1$. Under (3.3), $%
ROR(Y;w)=R_{11}(Y_{3}^{1};w)/R_{11}(Y_{3}^{0};w)$ in (3.2) becomes%
\begin{equation*}
\frac{P(Y_{3}^{1}=1|w,Q=1)/P(Y_{3}^{1}=0|w,Q=1)}{%
P(Y_{3}^{0}=1|w,Q=1)/P(Y_{3}^{0}=0|w,Q=1)}\simeq \frac{P(Y_{3}^{1}=1|w,Q=1)}{%
P(Y_{3}^{0}=1|w,Q=1)}\text{.}
\end{equation*}%
Hence, \textit{the proportional odds effect in (3.2) becomes the
proportional effect in (2.2):}%
\begin{equation*}
ROR(Y;w)-1\simeq \frac{E(Y_{3}^{1}-Y_{3}^{0}|w,Q=1)}{E(Y_{3}^{0}|w,Q=1)}%
\text{ \ \ \ \ under the rare event condition (3.3).}
\end{equation*}

$\qquad ROR(Y;w)$ can be estimated nonparametrically by substituting sample
analogs into the components of $ROR(Y;w)$. However, as was the case for DD
and $RR(w)$, this is not what practitioners would do. Instead, we apply
logistic regression next.

\subsection{Logit for Binary Response}

\qquad Consider the popular logistic binary choice panel data model for $%
Y_{it}^{d}$:%
\begin{equation}
Y_{t}^{d}=1[0<\beta _{t}+\beta _{q}Q+\beta _{d}d+W_{t}^{\prime }\beta
_{w}+U_{t}],\ \ \ \ \ U_{t}\sim Logistic\amalg (Q,W_{t}).  \tag{3.4}
\end{equation}%
This yields the RCS model for $Y^{d}=1[0<\beta _{2}+\beta _{\tau }S+\beta
_{q}Q+\beta _{d}d+W^{\prime }\beta _{w}+U]$, which then yields the logistic
RCS model for $Y=(1-D)Y^{0}+DY$ as in (2.4) to (2.5):%
\begin{eqnarray}
&&E(Y|W,Q,S)=(1-QS)\cdot E(Y^{0}|W,Q,S)+QS\cdot E(Y^{1}|W,Q,S)  \notag \\
&=&\frac{\exp (\beta _{2}+\beta _{\tau }S+\beta _{q}Q+\beta _{d}D+W^{\prime
}\beta _{w})}{1+\exp (\beta _{2}+\beta _{\tau }S+\beta _{q}Q+\beta
_{d}D+W^{\prime }\beta _{w})};  \TCItag{3.5}
\end{eqnarray}%
the last equality can be verified by substituting $D=QS=0,1$.

\qquad The logistic panel data model for $Y_{it}^{0}$ gives%
\begin{eqnarray}
R_{11}(Y_{3}^{0};w) &=&\exp (\beta _{2}+\beta _{\tau }+\beta _{q}+w^{\prime
}\beta _{w}),\text{ \ \ }R_{01}(Y_{3}^{0};w)=\exp (\beta _{2}+\beta _{\tau
}+w^{\prime }\beta _{w}),  \notag \\
R_{10}(Y_{2}^{0};w) &=&\exp (\beta _{2}+\beta _{q}+w^{\prime }\beta _{w}),%
\text{ \ \ \ \ \ \ \ \ \ }R_{00}(Y_{2}^{0};w)=\exp (\beta _{2}+w^{\prime
}\beta _{w}).  \TCItag{3.6}
\end{eqnarray}%
Hence, ID$_{ROR}$ holds for the logistic panel data model: due to (3.6),%
\begin{equation*}
ROR(Y^{0};w)=\left( \frac{R_{11}(Y_{3}^{0};w)}{R_{10}(Y_{2}^{0};w)}\right)
/\left( \frac{R_{01}(Y_{3}^{0};w)}{R_{00}(Y_{2}^{0};w)}\right) =1.
\end{equation*}%
Also, $R_{11}(Y_{3}^{1};w)=\exp (\beta _{2}+\beta _{\tau }+\beta _{q}+\beta
_{d}+w^{\prime }\beta _{w})$ and $R_{11}(Y_{3}^{0};w)$ in (3.6) give%
\begin{equation*}
ROR(Y;w)-1=\{R_{11}(Y_{3}^{1};w)/R_{11}(Y_{3}^{0};w)\}-1=\exp (\beta _{d})-1.
\end{equation*}%
Estimate $\beta _{d}$ with the logistic MLE with (3.5) to use $\exp (\beta
_{d})-1$ as the proportional odds effect on $Y$, which is also the
proportional effect when $Y=1$ is a rare event.

\qquad Suppose $\beta _{q\tau }tQ$ with $\beta _{q\tau }\neq 0$ appears as
an extra regressor in (3.4). Then the parallel trends do not hold for the
latent response $Y^{\ast }$. The appearance of $\beta _{q\tau }tQ$ also
ruins ID$_{ROR}$ for binary $Y$ because ID$_{ROR}$ becomes (2.7), just as $%
\beta _{q\tau }tQ$ ruins ID$_{RR}$ in (2.7). As in (2.7), using $tQ$ is an
easy way to test or allow for non-parallel trends in $Y^{\ast }$. The
comments made for (2.7) hold more or less the same for (3.4) and (3.5).

\qquad Suppose now that the slope of $d$ in (3.4) is $\beta _{d}(W_{t})$,
e.g., $\beta _{d}(W_{t})=\beta _{d0}+\beta _{dw}^{\prime }W_{t}$:%
\begin{equation*}
Y_{t}^{d}=1[0<\beta _{t}+\beta _{q}Q+\beta _{d}(W_{t})d+W_{t}^{\prime }\beta
_{w}+U_{t}].
\end{equation*}%
Then (3.5) and $ROR(Y;w)-1$ become, respectively,%
\begin{eqnarray*}
&&E(Y|W,Q,S)=\frac{\exp \{\beta _{2}+\beta _{\tau }S+\beta _{q}Q+\beta
_{d}(W)D+W^{\prime }\beta _{w}\}}{1+\exp \{\beta _{2}+\beta _{\tau }S+\beta
_{q}Q+\beta _{d}(W)D+W^{\prime }\beta _{w}\}}, \\
&&ROR(Y;w)-1=\exp \{\beta _{d}(w)\}-1.
\end{eqnarray*}

\subsection{Logit for Fractional Response}

\qquad When $Y$ takes on a value in $[0,1]$, $Y$ is a fractional response;
e.g., the proportion of asset invested in stocks. There are two types of
fractional response: (i) $P(Y=0$ or $Y=1)=0$ and (ii) $P(Y=0$ or $Y=1)>0$.
Since the logistic regression model (3.5) always gives a value in $(0,1)$,
the logistic regression can be adopted for type-(i) fractional response,
regardless of whether (3.5) is derived from some latent $Y^{\ast }$ or not.

\qquad As for type (ii), analogously to $Y=\max (0,Y^{\ast })$, we can use $%
Y=\max \{0,\min (Y^{\ast },1)\}$. Since the transformation $\max \{0,\min
(\cdot ,1)\}$ is not smooth, one may object to adopting (3.5) for type (ii).
However, as Santos Silva and Tenreyro (2011) justified adopting the
exponential regression for $\max (0,\cdot )$, Papke and Wooldridge (1996)
justified adopting the logistic regression for $\max \{0,\min (\cdot ,1)\}$.

\qquad Papke and Wooldridge maximize the logistic QMLE log-likelihood
function for $b$:%
\begin{equation*}
\sum_{i}\left\{ Y_{i}\ln \frac{\exp (X_{i}^{\prime }b)}{1+\exp
(X_{i}^{\prime }b)}+(1-Y_{i})\ln \frac{1}{1+\exp (X_{i}^{\prime }b)}\right\}
;
\end{equation*}%
$X_{i}$ and $b$ were defined for the Poisson QMLE.\ The first-order
condition is%
\begin{equation*}
\sum_{i}\left\{ Y_{i}-\frac{\exp (X_{i}^{\prime }b)}{1+\exp (X_{i}^{\prime
}b)}\right\} X_{i}=0\text{ \ \ \ \ (satisfied under (3.5)).}
\end{equation*}%
That is, the logistic QMLE applies to fractional response too, but as in
Poisson QMLE, a \textquotedblleft sandwich form\textquotedblright\
asymptotic variance estimator should be used. The maximum is unique, because
the second-order matrix $-\sum_{i}X_{i}X_{i}^{\prime }[\exp (X_{i}^{\prime
}b)/\{1+\exp (X_{i}^{\prime }b)\}^{2}]$ is n.d.

\section{Empirical Analysis}

\qquad In this section, we estimate the effects of the Affordable Care Act
Dependent Coverage Provision (`DCP') on various health outcomes. Under the
DCP that went into effect in September 2010, dependents can remain on the
parent's private health plan until age 26. The treatment group is dependents
aged 23-25, and the control group is dependents aged 27-29; 26 was excluded
due to the treatment status ambiguity.

\qquad Our data came from the Behavioral Risk Factor Surveillance System
(BRFSS) for years 2007-2013, which is health-related telephone surveys in
the U.S. Almost the same data were used in Barbaresco, Courtemanche and Qi
(2015) (`BCQ', henceforth), with small differences occurring due to updates,
imputed values, data cleaning, etc. As in BCQ, sampling weights are used in
estimation and cluster-robust standard errors are reported in the tables
below.

\qquad BCQ considered 18 outcomes, of which we use 12. Each outcome variable
has a different sample size, as we replaced \textquotedblleft Don't
Know\textquotedblright\ and \textquotedblleft Refused\textquotedblright\
with missing values. With the sample size in \{$\cdot $\}, the 12 outcome
variables are:\ \textbf{(1)} `any health insurance' \{127618\}, \textbf{(2)}%
\ `any primary (care) doctor' \{127533\}, \textbf{(3)}\ needed medical care
in past year not taken due to cost (`cost blocked care') \{108433\}, \textbf{%
(4)}\ current smoker \{126557\}, \textbf{(5)} `risky drinker (in past 30
days)' \{122035\}, \textbf{(6)}\ `obese (BMI$\geq $30)' \{121294\}, \textbf{%
(7)}\ `pregnant (while) unmarried' \{40006\}, \textbf{(8)}\ `(alcoholic)
drinks (in past) 30 days' \{121845\}, \textbf{(9)} BMI \{121290\}, \textbf{%
(10)} days of last 30 not in good mental health (`days poor mental')\
\{125681\}, \textbf{(11)}\ days of last 30 not in good physical health
(`days poor physical') \{125766\},\ and \textbf{(12)} days of last 30 with
health-related limitations (`days health limits') \{71079\}. The first seven
outcomes are binary, and the remaining five are non-negative (counts or
continuous).

\qquad Table 1 presents summary statistics on covariates: age, gender, race,
marital status, education, state unemployment rate, `any DCP' for whether
the state has any DCP mandate although the dependent may not be covered,
household income, the number of children, `cell phones only' (vs. cell phone
plus landline), student, and unemployed. Because the treatment group is
younger than control group by $2\sim 6$ years, the treatment group has fewer
married, fewer college degree, lower household income, fewer children, more
students, and more unemployed. Also, the treatment group has the lower state
unemployment rate, higher any DCP, and higher cell phone only.

\qquad Let `Lin-DD' stand for the usual linear model DD using (1.2). Table 2
shows the estimates for $\beta _{q\tau }$ (non-parallel trends) along with $%
\beta _{d}$ (treatment effect), although the effect of interest is the
proportional effect $\exp (\beta _{d})-1$. Poisson QMLE estimates are $%
\tilde{\beta}_{q\tau }$\ and $\tilde{\beta}_{d}$, whereas\ Lin-DD estimates
ignoring the LDV nature are $\hat{\beta}_{q\tau }$\ and $\hat{\beta}_{d}$.

\begin{center}
\begin{tabular}{cccccc}
\hline\hline
\multicolumn{6}{c}{Table 1. Summary Statistics of Covariates: Mean \&
Standard Deviation (SD)} \\ 
& {\small Treated} & {\small Control} &  & {\small Treated} & {\small Control%
} \\ 
\multicolumn{1}{l}{\small Covariates} & {\small Mean (SD)} & {\small Mean
(SD)} & \multicolumn{1}{l}{\small Covariates} & {\small Mean (SD)} & {\small %
Mean (SD)} \\ \hline
\multicolumn{3}{l}{\small Age (age 23 omitted)} & \multicolumn{3}{l}{\small %
Household income (less than \$10K omitted)} \\ 
{\small Age 24} & \multicolumn{1}{r}{\small 0.35 (0.48)} & \multicolumn{1}{r}%
{\small -} & {\small \$10K--\$15K} & \multicolumn{1}{r}{\small 0.07 (0.26)}
& \multicolumn{1}{r}{\small 0.05 (0.22)} \\ 
{\small Age 25} & \multicolumn{1}{r}{\small 0.32 (0.47)} & \multicolumn{1}{r}%
{\small -} & {\small \$15K--\$20K} & \multicolumn{1}{r}{\small 0.10 (0.30)}
& \multicolumn{1}{r}{\small 0.08 (0.27)} \\ 
{\small Age 27} & \multicolumn{1}{r}{\small -} & \multicolumn{1}{r}{\small %
0.31 (0.46)} & {\small \$20K--\$25K} & \multicolumn{1}{r}{\small 0.12 (0.32)}
& \multicolumn{1}{r}{\small 0.10 (0.30)} \\ 
{\small Age 28} & \multicolumn{1}{r}{\small -} & \multicolumn{1}{r}{\small %
0.35 (0.48)} & {\small \$25K--\$35K} & \multicolumn{1}{r}{\small 0.14 (0.35)}
& \multicolumn{1}{r}{\small 0.13 (0.34)} \\ 
{\small Age 29} & \multicolumn{1}{r}{\small -} & \multicolumn{1}{r}{\small %
0.35 (0.48)} & {\small \$35K--\$50K} & \multicolumn{1}{r}{\small 0.16 (0.37)}
& \multicolumn{1}{r}{\small 0.16 (0.37)} \\ 
\multicolumn{1}{l}{\small Female} & \multicolumn{1}{r}{\small 0.51 (0.50)} & 
\multicolumn{1}{r}{\small 0.51 (0.50)} & {\small \$50K--\$75K} & 
\multicolumn{1}{r}{\small 0.14 (0.35)} & \multicolumn{1}{r}{\small 0.19
(0.39)} \\ 
\multicolumn{3}{l}{\small Race (non-Hispanic whites omitted)} & {\small %
\$75K \$ over} & \multicolumn{1}{r}{\small 0.19 (0.39)} & \multicolumn{1}{r}%
{\small 0.24 (0.43)} \\ 
{\small Black} & \multicolumn{1}{r}{\small 0.11 (0.31)} & \multicolumn{1}{r}%
{\small 0.11 (0.32)} & \multicolumn{3}{l}{\small Number of children} \\ 
{\small Hispanic} & \multicolumn{1}{r}{\small 0.23 (0.42)} & 
\multicolumn{1}{r}{\small 0.22 (0.41)} & {\small 1} & \multicolumn{1}{r}%
{\small 0.23 (0.42)} & \multicolumn{1}{r}{\small 0.23 (0.42)} \\ 
{\small Others} & \multicolumn{1}{r}{\small 0.09 (0.28)} & \multicolumn{1}{r}%
{\small 0.08 (0.27)} & {\small 2} & \multicolumn{1}{r}{\small 0.16 (0.37)} & 
\multicolumn{1}{r}{\small 0.23 (0.42)} \\ 
\multicolumn{1}{l}{\small Married} & \multicolumn{1}{r}{\small 0.30 (0.46)}
& \multicolumn{1}{r}{\small 0.56 (0.50)} & {\small 3} & \multicolumn{1}{r}%
{\small 0.06 (0.23)} & \multicolumn{1}{r}{\small 0.11 (0.31)} \\ 
\multicolumn{3}{l}{\small Education (less than HS degree omitted)} & {\small %
4} & \multicolumn{1}{r}{\small 0.02 (0.13)} & \multicolumn{1}{r}{\small 0.04
(0.19)} \\ 
\multicolumn{1}{l}{\small High school (HS)} & \multicolumn{1}{r}{\small 0.28
(0.45)} & \multicolumn{1}{r}{\small 0.26 (0.44)} & {\small 5 or more} & 
\multicolumn{1}{r}{\small 0.01 (0.09)} & \multicolumn{1}{r}{\small 0.02
(0.12)} \\ 
{\small Non-4-yr coll.} & \multicolumn{1}{r}{\small 0.30 (0.49)} & 
\multicolumn{1}{r}{\small 0.27 (0.44)} &  & \multicolumn{1}{r}{} & 
\multicolumn{1}{r}{} \\ 
{\small Coll. graduate} & \multicolumn{1}{r}{\small 0.31 (0.46)} & 
\multicolumn{1}{r}{\small 0.37 (0.48)} & \multicolumn{1}{l}{\small Cell
phone only} & \multicolumn{1}{r}{\small 0.70 (0.46)} & \multicolumn{1}{r}%
{\small 0.67 (0.47)} \\ 
\multicolumn{1}{l}{\small State unemp. rate} & \multicolumn{1}{r}{\small %
7.09 (2.72)} & \multicolumn{1}{r}{\small 7.22 (2.73)} & \multicolumn{1}{l}%
{\small Student} & \multicolumn{1}{r}{\small 0.11 (0.31)} & 
\multicolumn{1}{r}{\small 0.05 (0.23)} \\ 
\multicolumn{1}{l}{\small Any DCP} & \multicolumn{1}{r}{\small 0.26 (0.44)}
& \multicolumn{1}{r}{\small 0.04 (0.20)} & \multicolumn{1}{l}{\small %
Unemployed} & \multicolumn{1}{r}{\small 0.13 (0.34)} & \multicolumn{1}{r}%
{\small 0.12 (0.32)} \\ \hline
\multicolumn{6}{c}{\small coll:\ college; Any DCP: the state has any DCP
mandate despite the person not covered} \\ \hline\hline
\end{tabular}
\end{center}

\qquad Three main findings emerge from Table 2, which are also seen in the
simulation part of the appendix:\ (i) RR and ROR estimates differ much from
Lin-DD estimates; (ii) the difference is overall greater for non-negative
responses than for binary responses, as all signs are the same for binary
responses but some signs differ for non-negative responses; and (iii) $\beta
_{q\tau }=0$ is rejected in Lin-DD more often than in RR and ROR. Comparing
RR, ROR and Lin-DD in their qualitative conclusions by testing for $\beta
_{d}=0$, they lead to the same qualitative conclusions, except for `drinks
30 days'.

\begin{center}
\begin{tabular}{ccccc}
\hline\hline
\multicolumn{5}{c}{Table 2. Non-Parallel Trends ($\beta _{q\tau }$) and
Treatment Effect ($\beta _{d}$): Estimate (t-value)} \\ 
& \multicolumn{2}{c}{RR and ROR} & \multicolumn{2}{c}{Lin-DD (Linear model
DD)} \\ 
\multicolumn{1}{l}{} & $\tilde{\beta}_{q\tau }$ & $\tilde{\beta}_{d}$ & DD $%
\hat{\beta}_{q\tau }$ & DD $\hat{\beta}_{d}$ \\ 
\multicolumn{1}{l}{Outcome variable} & Estimate (tv) & Estimate (tv) & 
Estimate (tv) & Estimate (tv) \\ \hline
\multicolumn{5}{c}{\textit{Binary response}} \\ 
\multicolumn{1}{l}{Any health insurance} & \multicolumn{1}{r}{-0.011 (-0.32)}
& \multicolumn{1}{r}{0.415 (3.13)} & \multicolumn{1}{r}{-0.002 (-0.37)} & 
\multicolumn{1}{r}{0.069 (2.84)} \\ 
\multicolumn{1}{l}{Any primary doctor} & \multicolumn{1}{r}{0.007 (0.28)} & 
\multicolumn{1}{r}{0.150 (1.32)} & \multicolumn{1}{r}{0.001 (0.23)} & 
\multicolumn{1}{r}{0.030 (1.22)} \\ 
\multicolumn{1}{l}{Cost blocked care} & \multicolumn{1}{r}{0.014 (1.56)} & 
\multicolumn{1}{r}{-0.167 (-1.64)} & \multicolumn{1}{r}{0.003 (2.12)} & 
\multicolumn{1}{r}{-0.029 (-1.79)} \\ 
\multicolumn{1}{l}{Current smoker} & \multicolumn{1}{r}{-0.051 (-2.57)} & 
\multicolumn{1}{r}{0.195 (2.98)} & \multicolumn{1}{r}{-0.009 (-2.74)} & 
\multicolumn{1}{r}{0.034 (3.23)} \\ 
\multicolumn{1}{l}{Risky drinker} & \multicolumn{1}{r}{0.032 (1.78)} & 
\multicolumn{1}{r}{-0.108 (-2.62)} & \multicolumn{1}{r}{0.006 (2.17)} & 
\multicolumn{1}{r}{-0.022 (-3.29)} \\ 
\multicolumn{1}{l}{Obese} & \multicolumn{1}{r}{0.033 (2.02)} & 
\multicolumn{1}{r}{-0.083 (-1.10)} & \multicolumn{1}{r}{0.007 (1.95)} & 
\multicolumn{1}{r}{-0.018 (-1.03)} \\ 
\multicolumn{1}{l}{Pregnant unmarried} & \multicolumn{1}{r}{0.008 (0.13)} & 
\multicolumn{1}{r}{-0.074 (-0.42)} & \multicolumn{1}{r}{0.000 (0.10)} & 
\multicolumn{1}{r}{-0.003 (-0.36)} \\ 
\multicolumn{5}{c}{\textit{Non-negative response}} \\ 
\multicolumn{1}{l}{Drinks 30 days} & \multicolumn{1}{r}{-0.037 (-1.22)} & 
\multicolumn{1}{r}{0.091 (0.76)} & \multicolumn{1}{r}{-1.420 (-4.69)} & 
\multicolumn{1}{r}{4.521 (3.19)} \\ 
\multicolumn{1}{l}{BMI} & \multicolumn{1}{r}{-0.002 (-1.45)} & 
\multicolumn{1}{r}{0.000 (-0.04)} & \multicolumn{1}{r}{-0.034 (-0.92)} & 
\multicolumn{1}{r}{-0.105 (-0.69)} \\ 
\multicolumn{1}{l}{Days poor mental} & \multicolumn{1}{r}{-0.006 (-0.23)} & 
\multicolumn{1}{r}{0.002 (0.01)} & \multicolumn{1}{r}{-0.016 (-0.15)} & 
\multicolumn{1}{r}{0.195 (0.41)} \\ 
\multicolumn{1}{l}{Days poor physical} & \multicolumn{1}{r}{0.008 (0.18)} & 
\multicolumn{1}{r}{-0.047 (-0.26)} & \multicolumn{1}{r}{-0.087 (-0.94)} & 
\multicolumn{1}{r}{0.429 (1.14)} \\ 
\multicolumn{1}{l}{Days health limits} & \multicolumn{1}{r}{-0.001 (-0.02)}
& \multicolumn{1}{r}{0.017 (0.11)} & \multicolumn{1}{r}{0.041 (0.35)} & 
\multicolumn{1}{r}{0.177 (0.35)} \\ \hline\hline
\end{tabular}
\end{center}

\qquad Turning to interpreting effect magnitude, proportional odds effects
are a little hard to interpret; e.g., DCP increases the odds ratio of `any
health insurance' by $42\%$. This should not be taken as a drastic effect,
because odds ratios can easily take on large values (and change much), which
is, in fact, one of the reasons why some researchers prefer ratios to
differences. Compared with the overall large magnitudes in proportional odds
effects for binary responses, the proportional effect magnitudes for
non-negative responses are in a much smaller scale and easy to interpret,
ranging just over $-0.047$ to $0.091$; e.g., DCP increases `drinks 30 days'
by $9.1\%$. As an example for proportional odds effects becoming
proportional effects for rare events, unmarried pregnancies are fairly rare (%
$4\sim 5\%$) in our data, and consequently, we can interpret the ROR
estimate $-0.074$ for `pregnant unmarried' as a $7\%$ decrease due to DCP.

\qquad BCQ checked out the parallel trend assumption with graphs plotting
the pre-treatment trends across the treatment and control groups. BCQ also
estimated their models using different time periods or using more aggregated
data. Whereas these are informal/indirect ways of testing for parallel
trends, our approach of using $tQ$ as an extra regressor provides a formal
test for parallel trends, as well as a simple way to allow for non-parallel
trends. The $\tilde{\beta}_{q\tau }$ estimates in Table 2 reveal that
parallel trend assumption in $Y^{\ast }$ and the analogous ID$_{RR}$/ID$%
_{ROR}$ assumption do not hold at least for `current smoker' and `obese',
and Lin-DD rejects $\beta _{q\tau }=0$ for even more outcomes.

\begin{center}
\begin{tabular}{ccc}
\hline\hline
\multicolumn{3}{c}{Table 3. Treatment Effects under Parallel Trends} \\ 
& RR and ROR & Lin-DD \\ 
\multicolumn{1}{l}{Outcome variable} & Estimate (tv) & Estimate (tv) \\ 
\hline
\multicolumn{3}{c}{\textit{Binary response}} \\ 
\multicolumn{1}{l}{Any health insurance} & \multicolumn{1}{r}{0.375 (4.55)}
& \multicolumn{1}{r}{0.061 (4.34)} \\ 
\multicolumn{1}{l}{Any primary doctor} & \multicolumn{1}{r}{0.176 (5.34)} & 
\multicolumn{1}{r}{0.034 (4.87)} \\ 
\multicolumn{1}{l}{Cost blocked care} & \multicolumn{1}{r}{-0.109 (-1.30)} & 
\multicolumn{1}{r}{-0.018 (-1.34)} \\ 
\multicolumn{1}{l}{Current smoker} & \multicolumn{1}{r}{0.009 (0.25)} & 
\multicolumn{1}{r}{0.001 (0.25)} \\ 
\multicolumn{1}{l}{Risky drinker} & \multicolumn{1}{r}{0.011 (0.38)} & 
\multicolumn{1}{r}{0.002 (0.31)} \\ 
\multicolumn{1}{l}{Obese} & \multicolumn{1}{r}{0.037 (1.17)} & 
\multicolumn{1}{r}{0.009 (1.22)} \\ 
\multicolumn{1}{l}{Pregnant unmarried} & \multicolumn{1}{r}{-0.045 (-0.44)}
& \multicolumn{1}{r}{-0.002 (-0.49)} \\ 
\multicolumn{3}{c}{\textit{Non-negative response}} \\ 
\multicolumn{1}{l}{Drinks 30 days} & \multicolumn{1}{r}{-0.043 (-0.58)} & 
\multicolumn{1}{r}{-0.708 (-0.49)} \\ 
\multicolumn{1}{l}{BMI} & \multicolumn{1}{r}{-0.008 (-2.52)} & 
\multicolumn{1}{r}{-0.228 (-3.15)} \\ 
\multicolumn{1}{l}{Days poor mental} & \multicolumn{1}{r}{-0.019 (-0.40)} & 
\multicolumn{1}{r}{0.137 (0.78)} \\ 
\multicolumn{1}{l}{Days poor physical} & \multicolumn{1}{r}{-0.019 (-0.43)}
& \multicolumn{1}{r}{0.109 (2.15)} \\ 
\multicolumn{1}{l}{Days health limits} & \multicolumn{1}{r}{0.014 (0.25)} & 
\multicolumn{1}{r}{0.327 (2.67)} \\ \hline\hline
\end{tabular}
\end{center}

\qquad To appreciate better how much difference allowing $\beta _{q\tau
}\neq 0$ makes, Table 3 repeats Table 2 under the restriction $\beta _{q\tau
}=0$ (i.e., without using $tQ$ as a regressor). The differences between
Tables 2 and 3 are huge both in terms of effect magnitude and t-value. In RR
and ROR, only `any health insurance' maintained its statistical
significance, whereas `current smoker' and `risky drinker' become
misleadingly insignificant by imposing $\beta _{q\tau }=0$ falsely. Also,
`any primary doctor' and BMI become significant by imposing $\beta _{q\tau
}=0$ unnecessarily. In Lin-DD as well, only `any health insurance' maintains
its statistical significance in Tables 2 and 3, whereas the statistical
significance of seven other outcomes is switched.

\qquad The main finding in BCQ is that DCP increases `any health insurance',
`any primary doctor' and `risky drinker', but decreases BMI. This finding is
similar to that of the RR and ROR column in Table 3, except for `risky
drinker' that is insignificant in Table 3. This similarity is due to $\beta
_{q\tau }=0$ assumed in both BCQ and Table 3.

\qquad Since Table 3 imposes the unnecessary restriction $\beta _{q\tau }=0$%
, it is interesting to compare the finding in BCQ to that in Table 2. The RR
and ROR column of Table 2 reveals significantly increasing effects on `any
health insurance' and `current smoker', and a significantly decreasing
effect on `risky drinker'. Hence, only the increasing effect on `any health
insurance' is shared by BCQ and the RR and ROR column of Table 2; the sign
of `risky drinker' changes across BCQ and the RR and ROR column of Table 2.
Overall, the differences due to allowing $\beta _{q\tau }\neq 0$ are large.

\section{Conclusions}

\qquad Difference in Differences (DD) is one of the most popular approaches
in finding the effect of a treatment $D$ on an outcome/response $Y$.
However, DD is suitable for linear models, and consequently, applying DD to
limited dependent variables (LDV's), or more generally to nonlinear models,
has been problematic. Many researchers with LDV's simply ignore the LDV
nature to use a linear model. The goal of this paper is to explore what can
be done in this case, and this paper obtained the following findings,
adopting the framework of generalized linear models with link functions.

\qquad First, when the LDV is a non-negative outcome such as count or
zero-censored response, `ratio in ratios (RR)' is more appropriate than DD,
because exponential regression models appear naturally in this context, and
RR removes the time and group effects to identify the treatment effect. The
semiparametric `Poisson Quasi-MLE' can be applied with $\beta _{d}D$ in the
model, and $\exp (\beta _{d})-1$ is the unit-free \textit{proportional
effect }$E(Y_{3}^{1}-Y_{3}^{0}|Q=1)/E(Y^{0}|Q=1)$, where $%
(Y_{3}^{0},Y_{3}^{1})$ are the potential outcomes in the post-treatment
period $3$, and $Q=1$ is the treatment-qualification dummy in DD.

\qquad Second, when the LDV is binary, fractional or multinomial, `ratio in
odds ratios (ROR)' is more appropriate than DD, because \textquotedblleft
normalized\textquotedblright\ exponential regression models appear naturally
in this context, and ROR removes the time and group effects to identify the
treatment effect. The binary/multinomial logit MLE can be applied with $%
\beta _{d}D$, and $\exp (\beta _{d})-1$ is the unit-free \textit{%
proportional odds effect}, which is not easy to interpret though, compared
with the proportional effect. Nevertheless, for rare events (i.e., $%
P(Y=0)\simeq 1$), the proportional odds effect becomes the proportional
effect. ROR is not applicable to ordinal responses, which however can be
reduced to binary responses in multiple ways, and then the overlapping
information in those multiple ways can be combined with minimum distance
estimation (see, e.g., Lee 2015).

\qquad Third, a simple interaction regressor $tQ$\ where $t$\ denotes time
allows testing for the critical DD, RR and ROR identification conditions
(for DD, the condition is called `parallel trends'). Namely, with $\beta
_{q\tau }$ being the slope of $tQ$, if $\beta _{q\tau }=0$, then the
identification conditions hold. Viewed differently, instead of testing for
the conditions, using $tQ$ as an extra regressor relaxes the identification
conditions for DD, RR and ROR.

\qquad Our empirical study, as well as the simulation study in the appendix,
revealed the importance of using RR or ROR instead of DD for LDV's. The
empirical study using as many as 12 outcome variables showed that RR and ROR
give much different findings from DD. Also, using $\beta _{q\tau }tQ$ made
big differences in empirical findings, compared with imposing the
parallel-trend-type restriction $\beta _{q\tau }=0$ unnecessarily.\bigskip

\begin{center}
{\LARGE APPENDIX}
\end{center}

\textbf{Simulation Study}\bigskip

\qquad Our simulation study addresses four LDV models: (i) positive
continuous response, (ii) count response, (iii) zero-censored response with
many zeros, and (iv) binary response. Poisson QMLE is applied to (i), (ii)
and (iii), and logistic MLE to (iv); their estimates are compared with the
usual linear model DD (`Lin-DD'). Fractional response is not tried because
it is not yet clear how to generate fractional responses subject to the
exponential regression model, and multinomial response is addressed
separately in the next section because it is inconceivable to apply Lin-DD
to multinomial response.

\qquad In the following, we explain (i) and Table A1 in detail, from which
it will be clear how (ii), (iii) and (iv) are dealt with and how to
interpret the other tables. In all cases, the effect of interest is $\exp
(\beta _{d})-1$, which is the proportional (odds) effect, but we take $\beta
_{d}$ as the effect of interest because knowing $\beta _{d}$ is equivalent
to knowing $\exp (\beta _{d})-1$.

\qquad For (i) positive continuous response, we generate $Y_{it}$ for $%
t=0,1,2,3$:%
\begin{eqnarray}
Y_{it} &=&\exp (\beta _{t}+\beta _{q}Q_{i}+\beta _{q\tau }tQ_{i}+\beta
_{d}D_{it}+U_{it})\text{ \ \ \ \ where}  \TCItag{A.1} \\
P(Q_{i} &=&0)=P(Q_{i}=1)=0.5\text{, \ }D_{it}=Q_{i}1[t=3]\text{, \ }%
U_{i0},U_{i1},U_{i2},U_{i3}\text{ iid }N(0,1),  \notag \\
\beta _{0} &=&-2,\ \beta _{1}=-2,\ \beta _{2}=-1,\ \beta _{3}=-1\text{, \ }%
\beta _{q}=0.5,\ \ \beta _{q\tau }=0,\ 0.5,\ \ \beta _{d}=0,\ 0.5;  \notag
\end{eqnarray}%
recalling (2.7), $\beta _{q\tau }=0$ makes the parallel trends hold in $%
Y^{\ast }$ and ID$_{RR}$ hold in $Y$, but $\beta _{q\tau }=0.5$ violates
both. The simulation design is somewhat sensitive to the parameter values,
as the exponential function can \textquotedblleft blow up\textquotedblright\
sometimes to make Poisson QMLE fail to converge. The parameter values in
(A.1) are chosen to avoid this pitfall.

\qquad From the $Y_{it}$'s in (A.1), the RCS response $Y_{i}$ and its
regressor $X_{i}$ are obtained:%
\begin{eqnarray}
&&S_{i}\text{ is the sampled period for }i\text{, \ }S_{it}\equiv 1[S_{i}=t]%
\text{, \ }P(S_{it}=1)=0.25\text{ for all }t\text{,}  \notag \\
&&Y_{i}=\sum_{t=0}^{3}S_{it}Y_{it},\ \ \ Q_{i}^{\tau }\equiv
Q_{i}\sum_{t=1}^{3}S_{it}t,\ \ \ X_{i}\equiv
(1,S_{i1},S_{i2},S_{i3},Q_{i},Q_{i}^{\tau },D_{i})^{\prime }  \TCItag{A.2} \\
&&\text{for parameters \ \ }\{\beta _{0}+\ln (1.64),\ \beta _{1}-\beta
_{0},\ \beta _{2}-\beta _{0},\ \beta _{3}-\beta _{0},\ \beta _{q},\beta
_{q\tau },\beta _{d}\};  \notag
\end{eqnarray}%
$1.64$ comes from $E\{\exp (U_{it})\}=1.64$ with $U_{it}\sim N(0,1)$, which
appears due to%
\begin{equation*}
E(Y_{it}|Q_{i},S_{i})=E\{\exp (\beta _{t}+\beta _{q}Q_{i}+\beta _{q\tau
}tQ_{i}+\beta _{d}D_{it})\cdot E\{\exp (U_{it})\}.
\end{equation*}%
For the other LDV models, the RCS data are generated analogously.

\qquad Table A1 presents the simulation results with $5,000$ repetitions for 
$N=250$ and $10,000$, where each entry consists of the absolute bias (%
\TEXTsymbol{\vert}Bias\TEXTsymbol{\vert}), SD, and Root Mean Squared Error
(RMSE) for the $\beta _{q\tau }$ and $\beta _{d}$ estimates.

\qquad With $N=250$, Lin-DD estimates $\hat{\beta}_{q\tau }$ and $\hat{\beta}%
_{d}$ do sometimes better than the Poisson QMLE $\tilde{\beta}_{q\tau }$ and 
$\tilde{\beta}_{d}$, but this is due to the low SD's; the \TEXTsymbol{\vert}%
Bias\TEXTsymbol{\vert} of Lin-DD $\hat{\beta}_{d}$ is huge in several cases.
With $N=10,000$, the \TEXTsymbol{\vert}Bias\TEXTsymbol{\vert}'s for the
Lin-DD estimates remain almost the same as those with $N=250$ whereas the
gaps in SD between Lin-DD and Poisson QMLE are reduced, and consequently,
Poisson QMLE does better than Lin-DD. Using $tQ$ solves the problem of ID$%
_{RR}$ violation for Poisson QMLE, but not for Lin-DD; $\hat{\beta}_{q\tau }$
in Lin-DD is biased much even when $\beta _{q\tau }=0$. In short, Table A1
demonstrates that Lin-DD is highly biased when the true model is exponential
for positive $Y$.

\begin{center}
\begin{tabular}{cllll}
\hline\hline
\multicolumn{5}{c}{Table A1. Positive $Y$: \TEXTsymbol{\vert}Bias\TEXTsymbol{%
\vert}, SD and (RMSE)} \\ 
$N=250$ & $\beta _{q\tau },\beta _{d}$:$\ 0,\ 0$ & $\beta _{q\tau },\beta
_{d}$:$\ 0.5,\ 0$ & $\beta _{q\tau },\beta _{d}$:$\ 0,\ 0.5$ & $\beta
_{q\tau },\beta _{d}$:$\ 0.5,\ 0.5$ \\ \hline
$\tilde{\beta}_{q\tau }$ & \multicolumn{1}{r}{0.00 0.23 (0.23)} & 
\multicolumn{1}{r}{0.00 0.24 (0.24)} & \multicolumn{1}{r}{0.00 0.23 (0.23)}
& \multicolumn{1}{r}{0.00 0.24 (0.24)} \\ 
$\tilde{\beta}_{d}$ & \multicolumn{1}{r}{0.01 0.60 (0.60)} & 
\multicolumn{1}{r}{0.01 0.60 (0.60)} & \multicolumn{1}{r}{0.01 0.60 (0.60)}
& \multicolumn{1}{r}{0.01 0.60 (0.60)} \\ 
DD $\hat{\beta}_{q\tau }$ & \multicolumn{1}{r}{0.12 0.15 (0.19)} & 
\multicolumn{1}{r}{0.48 0.33 (0.59)} & \multicolumn{1}{r}{0.12 0.15 (0.19)}
& \multicolumn{1}{r}{0.48 0.33 (0.59)} \\ 
DD $\hat{\beta}_{d}$ & \multicolumn{1}{r}{0.08 0.46 (0.47)} & 
\multicolumn{1}{r}{1.05 1.38 (1.73)} & \multicolumn{1}{r}{0.08 0.56 (0.57)}
& \multicolumn{1}{r}{3.48 1.96 (4.00)} \\ 
$N=10000$ &  &  &  &  \\ 
$\tilde{\beta}_{q\tau }$ & \multicolumn{1}{r}{0.00 0.04 (0.04)} & 
\multicolumn{1}{r}{0.00 0.04 (0.04)} & \multicolumn{1}{r}{0.00 0.04 (0.04)}
& \multicolumn{1}{r}{0.00 0.04 (0.04)} \\ 
$\tilde{\beta}_{d}$ & \multicolumn{1}{r}{0.00 0.10 (0.10)} & 
\multicolumn{1}{r}{0.00 0.10 (0.10)} & \multicolumn{1}{r}{0.00 0.10 (0.10)}
& \multicolumn{1}{r}{0.00 0.10 (0.10)} \\ 
DD $\hat{\beta}_{q\tau }$ & \multicolumn{1}{r}{0.12 0.02 (0.13)} & 
\multicolumn{1}{r}{0.48 0.05 (0.49)} & \multicolumn{1}{r}{0.12 0.02 (0.13)}
& \multicolumn{1}{r}{0.48 0.05 (0.49)} \\ 
DD $\hat{\beta}_{d}$ & \multicolumn{1}{r}{0.08 0.07 (0.11)} & 
\multicolumn{1}{r}{1.03 0.22 (1.05)} & \multicolumn{1}{r}{0.07 0.09 (0.11)}
& \multicolumn{1}{r}{3.44 0.30 (3.45)} \\ \hline
\multicolumn{5}{l}{$\beta _{q\tau }=0$ for parallel trends in $Y^{\ast }$ \&
ID$_{RR}$ in $Y$; $\ \beta _{d}$ ($\exp (\beta _{d})-1$) is the desired} \\ 
\multicolumn{5}{l}{effect; $\ \tilde{\beta}_{q\tau }$,$\ \tilde{\beta}_{d}$%
:\ Poisson QMLE;$\ \ \hat{\beta}_{q\tau }$,$\ \hat{\beta}_{d}$: linear-model
DD.} \\ \hline\hline
\end{tabular}
\end{center}

\qquad For (ii) count response, similarly to (A.1), $Y_{it}$ is generated
from the Poisson distribution with parameter $\exp (\beta _{t}+\beta
_{q}Q_{i}+\beta _{q\tau }tQ_{i}+\beta _{d}D_{it})$ for $t=0,1,2,3$. Then $%
Y_{i}$ and $X_{i}$ are generated as in (A.2), and Poisson QMLE is
implemented, which is actually the Poisson MLE. The same parameters as in
(A.2) are estimated except for the intercept because $\ln (1.64)$ is no more
present. Table A2 presents the simulation results, and what was mentioned
for Table A1 applies to Table A2 almost word to word.

\qquad For (iii) zero-censored response, we use (2.9) where $M_{i}\sim
Poisson(1)$ with $P(M_{i}=0)=0.37$ and $Y_{it}=\sum_{j=0}^{M_{i}}Z_{ijt}$
with\ $Z_{ijt}=\exp \{\beta _{t}+\beta _{q}Q_{i}+\beta _{q\tau }tQ_{i}+\beta
_{d}D_{it}+N(0,1)\}$. The same parameters as in (A.2) are estimated except
for the intercept because $\exp (1)$ from $E(M)$ is added to $\beta _{0}$ in
view of (2.9). Despite the big difference in the data generating processes,
Table A3 differ little from Tables A1 and A2, and all comments made for
Tables A1 and A2 apply to Table A3 as well. The similarities in the findings
from Tables A1-A3 seem to stem from the common exponential regression
specification.

\begin{center}
\begin{tabular}{cllll}
\hline\hline
\multicolumn{5}{c}{Table A2. Poisson Count $Y$: \TEXTsymbol{\vert}Bias%
\TEXTsymbol{\vert}, SD and (RMSE)} \\ 
$N=250$ & $\beta _{q\tau },\beta _{d}$:$\ 0,\ 0$ & $\beta _{q\tau },\beta
_{d}$:$\ 0.5,\ 0$ & $\beta _{q\tau },\beta _{d}$:$\ 0,\ 0.5$ & $\beta
_{q\tau },\beta _{d}$:$\ 0.5,\ 0.5$ \\ \hline
$\tilde{\beta}_{q\tau }$ & \multicolumn{1}{r}{0.01 0.40 (0.40)} & 
\multicolumn{1}{r}{0.00 0.38 (0.38)} & \multicolumn{1}{r}{0.01 0.40 (0.40)}
& \multicolumn{1}{r}{0.00 0.38 (0.38)} \\ 
$\tilde{\beta}_{d}$ & \multicolumn{1}{r}{0.00 0.85 (0.85)} & 
\multicolumn{1}{r}{0.00 0.78 (0.78)} & \multicolumn{1}{r}{0.01 0.84 (0.84)}
& \multicolumn{1}{r}{0.00 0.78 (0.78)} \\ 
DD $\hat{\beta}_{q\tau }$ & \multicolumn{1}{r}{0.08 0.10 (0.13)} & 
\multicolumn{1}{r}{0.10 0.14 (0.17)} & \multicolumn{1}{r}{0.08 0.10 (0.13)}
& \multicolumn{1}{r}{0.10 0.14 (0.17)} \\ 
DD $\hat{\beta}_{d}$ & \multicolumn{1}{r}{0.06 0.31 (0.31)} & 
\multicolumn{1}{r}{0.61 0.47 (0.77)} & \multicolumn{1}{r}{0.17 0.33 (0.37)}
& \multicolumn{1}{r}{1.88 0.53 (1.95)} \\ 
$N=10000$ &  &  &  &  \\ 
$\tilde{\beta}_{q\tau }$ & \multicolumn{1}{r}{0.00 0.06 (0.06)} & 
\multicolumn{1}{r}{0.00 0.05 (0.05)} & \multicolumn{1}{r}{0.00 0.06 (0.06)}
& \multicolumn{1}{r}{0.00 0.05 (0.05)} \\ 
$\tilde{\beta}_{d}$ & \multicolumn{1}{r}{0.00 0.12 (0.12)} & 
\multicolumn{1}{r}{0.00 0.10 (0.10)} & \multicolumn{1}{r}{0.00 0.11 (0.11)}
& \multicolumn{1}{r}{0.00 0.10 (0.10)} \\ 
DD $\hat{\beta}_{q\tau }$ & \multicolumn{1}{r}{0.08 0.02 (0.08)} & 
\multicolumn{1}{r}{0.10 0.02 (0.10)} & \multicolumn{1}{r}{0.08 0.02 (0.08)}
& \multicolumn{1}{r}{0.10 0.02 (0.10)} \\ 
DD $\hat{\beta}_{d}$ & \multicolumn{1}{r}{0.05 0.05 (0.07)} & 
\multicolumn{1}{r}{0.62 0.07 (0.63)} & \multicolumn{1}{r}{0.16 0.05 (0.16)}
& \multicolumn{1}{r}{1.89 0.08 (1.89)} \\ \hline
\multicolumn{5}{l}{$\beta _{q\tau }=0$ for parallel trends in $Y^{\ast }$ \&
ID$_{RR}$ in $Y$; $\ \beta _{d}$ ($\exp (\beta _{d})-1$) is the desired} \\ 
\multicolumn{5}{l}{effect; $\ \tilde{\beta}_{q\tau }$,$\ \tilde{\beta}_{d}$%
:\ Poisson QMLE;$\ \ \hat{\beta}_{q\tau }$,$\ \hat{\beta}_{d}$: linear-model
DD.} \\ \hline\hline
\end{tabular}%
$\bigskip $

\begin{tabular}{cllll}
\hline\hline
\multicolumn{5}{c}{Table A3. Zero-Censored $Y$: \TEXTsymbol{\vert}Bias%
\TEXTsymbol{\vert}, SD and (RMSE)} \\ 
$N=250$ & $\beta _{q\tau },\beta _{d}$:$\ 0,\ 0$ & $\beta _{q\tau },\beta
_{d}$:$\ 0.5,\ 0$ & $\beta _{q\tau },\beta _{d}$:$\ 0,\ 0.5$ & $\beta
_{q\tau },\beta _{d}$:$\ 0.5,\ 0.5$ \\ \hline
$\tilde{\beta}_{q\tau }$ & \multicolumn{1}{r}{0.01 0.31 (0.31)} & 
\multicolumn{1}{r}{0.01 0.32 (0.32)} & \multicolumn{1}{r}{0.01 0.31 (0.31)}
& \multicolumn{1}{r}{0.01 0.32 (0.32)} \\ 
$\tilde{\beta}_{d}$ & \multicolumn{1}{r}{0.01 0.80 (0.80)} & 
\multicolumn{1}{r}{0.02 0.80 (0.80)} & \multicolumn{1}{r}{0.01 0.80 (0.80)}
& \multicolumn{1}{r}{0.02 0.80 (0.80)} \\ 
DD $\hat{\beta}_{q\tau }$ & \multicolumn{1}{r}{0.12 0.19 (0.22)} & 
\multicolumn{1}{r}{0.47 0.42 (0.63)} & \multicolumn{1}{r}{0.12 0.19 (0.22)}
& \multicolumn{1}{r}{0.47 0.42 (0.63)} \\ 
DD $\hat{\beta}_{d}$ & \multicolumn{1}{r}{0.07 0.59 (0.60)} & 
\multicolumn{1}{r}{1.04 1.76 (2.04)} & \multicolumn{1}{r}{0.07 0.71 (0.72)}
& \multicolumn{1}{r}{3.44 2.48 (4.24)} \\ 
$N=10000$ &  &  &  &  \\ 
$\tilde{\beta}_{q\tau }$ & \multicolumn{1}{r}{0.00 0.05 (0.05)} & 
\multicolumn{1}{r}{0.00 0.05 (0.05)} & \multicolumn{1}{r}{0.00 0.05 (0.05)}
& \multicolumn{1}{r}{0.00 0.05 (0.05)} \\ 
$\tilde{\beta}_{d}$ & \multicolumn{1}{r}{0.00 0.12 (0.12)} & 
\multicolumn{1}{r}{0.00 0.13 (0.13)} & \multicolumn{1}{r}{0.00 0.12 (0.12)}
& \multicolumn{1}{r}{0.00 0.13 (0.13)} \\ 
DD $\hat{\beta}_{q\tau }$ & \multicolumn{1}{r}{0.12 0.03 (0.13)} & 
\multicolumn{1}{r}{0.48 0.07 (0.49)} & \multicolumn{1}{r}{0.12 0.03 (0.13)}
& \multicolumn{1}{r}{0.48 0.07 (0.49)} \\ 
DD $\hat{\beta}_{d}$ & \multicolumn{1}{r}{0.08 0.09 (0.12)} & 
\multicolumn{1}{r}{1.03 0.28 (1.06)} & \multicolumn{1}{r}{0.07 0.11 (0.13)}
& \multicolumn{1}{r}{3.43 0.39 (3.46)} \\ \hline
\multicolumn{5}{l}{$\beta _{q\tau }=0$ for parallel trends in $Y^{\ast }$ \&
ID$_{RR}$ in $Y$; $\ \beta _{d}$ ($\exp (\beta _{d})-1$) is the desired} \\ 
\multicolumn{5}{l}{effect; $\tilde{\beta}_{q\tau }$,$\tilde{\beta}_{d}$:\
Poisson QMLE;$\ \ \hat{\beta}_{q\tau }$,$\hat{\beta}_{d}$: linear model DD.}
\\ \hline\hline
\end{tabular}
\end{center}

\qquad For (iv) binary response, $Y_{it}$ is generated with Logistic error $%
U_{it}$:%
\begin{equation*}
Y_{it}=1[0<\beta _{t}+\beta _{q}Q_{i}+\beta _{q\tau }tQ_{i}+\beta
_{d}D_{it}+U_{it}],\ \ \ U_{i0},U_{i1},U_{i2},U_{i3}\text{ are iid Logistic}.
\end{equation*}%
`$\beta _{q\tau }=0$' makes the parallel trends hold in term of $Y^{\ast }$,
and makes ID$_{ROR}$ hold in terms of $Y$; $\beta _{q\tau }=0.5$ violates
both of these.

\begin{center}
\begin{tabular}{cllll}
\hline\hline
\multicolumn{5}{c}{Table A4. Binary $Y$: \TEXTsymbol{\vert}Bias\TEXTsymbol{%
\vert}, SD and (RMSE)} \\ 
$N=250$ & $\beta _{q\tau },\beta _{d}$:$\ 0,\ 0$ & $\beta _{q\tau },\beta
_{d}$:$\ 0.5,\ 0$ & $\beta _{q\tau },\beta _{d}$:$\ 0,\ 0.5$ & $\beta
_{q\tau },\beta _{d}$:$\ 0.5,\ 0.5$ \\ \hline
$\tilde{\beta}_{q\tau }$ & 0.01 0.50 (0.50) & 0.01 0.51 (0.51) & 0.01 0.50
(0.50) & 0.01 0.51 (0.51) \\ 
$\tilde{\beta}_{d}$ & 0.02 1.16 (1.16) & 0.04 1.18 (1.19) & 0.04 1.15 (1.15)
& 0.13 1.54 (1.55) \\ 
DD $\hat{\beta}_{q\tau }$ & 0.02 0.07 (0.08) & 0.35 0.08 (0.36) & 0.02 0.07
(0.08) & 0.35 0.08 (0.36) \\ 
DD $\hat{\beta}_{d}$ & 0.02 0.21 (0.21) & 0.02 0.21 (0.21) & 0.39 0.21 (0.45)
& 0.43 0.20 (0.48) \\ 
$N=10000$ &  &  &  &  \\ 
$\tilde{\beta}_{q\tau }$ & 0.00 0.07 (0.07) & 0.00 0.07 (0.07) & 0.00 0.07
(0.07) & 0.00 0.07 (0.07) \\ 
$\tilde{\beta}_{d}$ & 0.00 0.17 (0.17) & 0.00 0.17 (0.17) & 0.00 0.17 (0.17)
& 0.00 0.17 (0.17) \\ 
DD $\hat{\beta}_{q\tau }$ & 0.02 0.01 (0.03) & 0.35 0.01 (0.35) & 0.02 0.01
(0.03) & 0.35 0.01 (0.35) \\ 
DD $\hat{\beta}_{d}$ & 0.02 0.03 (0.04) & 0.02 0.03 (0.04) & 0.39 0.03 (0.39)
& 0.43 0.03 (0.43) \\ \hline
\multicolumn{5}{l}{$\beta _{q\tau }=0$ for parallel trends in $Y^{\ast }$ \&
ID$_{ROR}$ in $Y$; $\ \beta _{d}$ ($\exp (\beta _{d})-1$) is the desired} \\ 
\multicolumn{5}{l}{effect; $\ \tilde{\beta}_{q\tau }$,$\tilde{\beta}_{d}$:\
logit estimates;$\ \ \hat{\beta}_{q\tau }$,$\hat{\beta}_{d}$: linear-model
DD.} \\ \hline\hline
\end{tabular}
\end{center}

\qquad Table A4 addresses binary $Y$. Since the logistic regression is used
in Table A4 instead of the exponential regression in Tables A1-A3, the
results in Table A4 differ much from those in Tables A1-A3. First, the
overall magnitude of \TEXTsymbol{\vert}Bias\TEXTsymbol{\vert} is much
smaller than in Tables A1-A3. Second, surprisingly, when $\beta _{q\tau
}=\beta _{d}=0$, the Lin-DD estimates with almost zero bias do several times
better than the logistic MLE estimates. Third, biases in Lin-DD are
persistent even when $N$ increases to $10000$, which implies that Lin-DD
will be eventually dominated by logistic MLE for a large enough $N$.
Nevertheless, less harm is seen in using Lin-DD for binary response,
compared with the other LDV's.\bigskip 

\textbf{Multinomial Logit for DD with Multinomial Response\bigskip }

\textit{Identification\bigskip }

\qquad For multinomial response $Y$ taking on a value among $0,1,...,C$
classes, define the `class-$c$ odds' (with the base class $0$) conditional
on $(W=w,Q=q,S=s)$ as%
\begin{eqnarray*}
&&\ R_{qs}^{c}(Y;w)\equiv \frac{P(Y=c|w,Q=q,S=s)}{P(Y=0|w,Q=q,S=s)}\text{ \
\ \ \ \ \ which implies} \\
&&\ R_{11}^{c}(Y;w)=R_{11}^{c}(Y_{3}^{1};w),\text{ \ \ \ \ }%
R_{01}^{c}(Y;w)=R_{01}^{c}(Y_{3}^{0};w), \\
&&\ R_{10}^{c}(Y;w)=R_{10}^{c}(Y_{2}^{0};w),\ \ \ \ \
R_{00}^{c}(Y;w)=R_{00}^{c}(Y_{2}^{0};w),
\end{eqnarray*}%
analogously to (3.1). Also define `class-$c$ ROR conditional on $W=w$':%
\begin{equation*}
ROR^{c}(Y;w)\equiv \left( \frac{R_{11}^{c}(Y;w)}{R_{10}^{c}(Y;w)}\right)
/\left( \frac{R_{01}^{c}(Y;w)}{R_{00}^{c}(Y;w)}\right) .
\end{equation*}%
The identification condition for $ROR^{c}$ with multinomial response is%
\begin{equation}
ROR^{c}(Y^{0};w)=\left( \frac{R_{11}^{c}(Y_{3}^{0};w)}{%
R_{10}^{c}(Y_{2}^{0};w)}\right) /\left( \frac{R_{01}^{c}(Y_{3}^{0};w)}{%
R_{00}^{c}(Y_{2}^{0};w)}\right) =1.  \tag{ID$_{RORc}$}
\end{equation}

\qquad As in (3.2), $ROR^{c}(Y;w)-1$ is equal to the `class-$c$ proportional
odds effect on the treated at the post-treatment period $t=3$':%
\begin{equation*}
ROR^{c}(Y;w)-1=\frac{R_{11}^{c}(Y_{3}^{1};w)-R_{11}^{c}(Y_{3}^{0};w)}{%
R_{11}^{c}(Y_{3}^{0};w)}\ \ \ \ \ \ \ \ \text{under ID}_{RORc}.
\end{equation*}%
Also, as in (3.3), if $Y=c\neq 0$ is a rare event in the sense of (3.3), then%
\begin{equation}
ROR^{c}(Y;w)-1\simeq \frac{P(Y_{3}^{1}=c|w,Q=1)-P(Y_{3}^{0}=c|w,Q=1)}{%
P(Y_{3}^{0}=c|w,Q=1)}  \tag{A.3}
\end{equation}%
which is the class-$c$ proportional effect on the treated at the
post-treatment period.\bigskip

\textit{Estimation\bigskip }

\qquad In panel multinomial choice with classes $c=0,1,...,C$, there are a
few possibilities for regressors, depending on whether they vary across
subjects, classes or times. Here, we consider three types of regressors:\ $%
A_{i}$ varying only across subjects (e.g., race), $H_{it}$ varying only
across subjects and times (e.g., income), and $W_{ict}$ varying across
subjects, classes and times (e.g., expense from choosing class $c$\textbf{).}
Let the `latent utility from class $c$' of subject $i$ at period $t=2,3$ be%
\begin{equation}
L_{ict}^{d}\equiv \beta _{tc}+\beta _{qc}Q_{i}+\beta _{dc}d+\beta
_{ac}^{\prime }A_{i}+\beta _{hc}^{\prime }H_{it}+\beta _{wc}^{\prime
}W_{ict}+U_{ict},\ \ \ c=0,1,...,C  \tag{A.4}
\end{equation}%
where the error terms $(U_{i02},...,U_{iC2},\ U_{i03},...,U_{iC3})$ are iid
with the type-I extreme value distribution, and independent of all
regressors at all times (`strict exogeneity').

\qquad The potential choice $Y_{it}^{d}$ with $D=d$ is%
\begin{equation*}
Y_{it}^{d}=\sum_{j=0}^{C}(j\times 1[L_{ijt}^{d}>L_{ikt}^{d}\text{ for all }%
k\neq j])\text{;}
\end{equation*}%
$Y_{it}^{d}$ takes on $0,1,...,C$, depending on which class gives the
maximum utility. Using (A.4), the choice probabilities for the untreated
responses $Y_{it}^{0}=0,1,...,C$ are:%
\begin{eqnarray*}
&&P(Y_{it}^{0}=c|Q_{i},A_{i},H_{it},W_{i0t},...,W_{iCt}) \\
&=&\frac{\exp (\beta _{tc}+\beta _{qc}Q_{i}+\beta _{ac}^{\prime }A_{i}+\beta
_{hc}^{\prime }H_{it}+\beta _{wc}^{\prime }W_{ict})}{\sum_{j=0}^{C}\exp
(\beta _{tj}+\beta _{qj}Q_{i}+\beta _{aj}^{\prime }A_{i}+\beta _{hj}^{\prime
}H_{it}+\beta _{wj}^{\prime }W_{ijt})} \\
\  &=&\frac{\exp (\Delta \beta _{tc}+\Delta \beta _{qc}Q_{i}+\Delta \beta
_{ac}^{\prime }A_{i}+\Delta \beta _{hc}^{\prime }H_{it}-\beta _{w0}^{\prime
}W_{i0t}+\beta _{wc}^{\prime }W_{ict})}{1+\sum_{j=1}^{C}\exp (\Delta \beta
_{tj}+\Delta \beta _{qj}Q_{i}+\Delta \beta _{aj}^{\prime }A_{i}+\Delta \beta
_{hj}^{\prime }H_{it}-\beta _{w0}^{\prime }W_{i0t}+\beta _{wj}^{\prime
}W_{ijt})}, \\
&&\Delta \beta _{tj}\equiv \beta _{tj}-\beta _{t0},\ \ \Delta \beta
_{qj}\equiv \beta _{qj}-\beta _{q0},\ \ \Delta \beta _{aj}\equiv \beta
_{aj}-\beta _{a0},\ \ \Delta \beta _{hj}\equiv \beta _{hj}-\beta _{h0};
\end{eqnarray*}

the second equality holds, dividing through by $\exp (\beta _{t0}+\beta
_{q0}Q_{i}+\beta _{a0}^{\prime }A_{i}+\beta _{h0}^{\prime }H_{it}+\beta
_{w0}^{\prime }W_{i0t})$ for the base class $c=0$. The numerator of the last
ratio becomes one for the base class. ID$_{RORc}$ holds for $%
P(Y_{it}^{0}=c|\cdot )$, analogously to the proof for ID$_{ROR}$.

\qquad Analogously derive the model for $P(Y_{it}^{1}=c|\cdot )$, which then
gives ($i$ omitted)%
\begin{eqnarray*}
&&R_{11}^{c}(Y^{d};w)=\exp (\Delta \beta _{3c}+\Delta \beta _{qc}Q+\Delta
\beta _{dc}d+\Delta \beta _{ac}^{\prime }A+\Delta \beta _{hc}^{\prime
}H_{3}-\beta _{w0}^{\prime }W_{03}+\beta _{wc}^{\prime }W_{c3}) \\
&&\ \ \text{where \ \ \ \ }\Delta \beta _{dc}\equiv \beta _{dc}-\beta _{d0}.
\end{eqnarray*}%
Since $P(Y_{it}^{1}=c|\cdot )$ differs from $P(Y_{it}^{0}=c|\cdot )$ only in
the extra term $\Delta \beta _{dc}$, we get the class-$c$ proportional
effect under the rare event condition (3.3):%
\begin{equation*}
\{R_{11}^{c}(Y_{3}^{1};w)/R_{11}^{c}(Y_{3}^{0};w)\}-1=\exp (\Delta \beta
_{dc})-1.
\end{equation*}

\qquad In RCS, omitting the subscript $i$, we observe $Y\equiv
(1-S)Y_{2}+SY_{3}$ where $Y_{t}\ (=0,...,C)$ is the realized choice at $t$,
along with $Q$,$\ S$ and%
\begin{equation*}
A,\ H\equiv (1-S)H_{2}+SH_{3},\ \ W_{0}\equiv (1-S)W_{02}+SW_{03},\ ...,\
W_{C}\equiv (1-S)W_{C2}+SW_{C3}.
\end{equation*}

The RCS choice probabilities are, with $\Delta \beta _{\tau j}\equiv \Delta
\beta _{3j}-\Delta \beta _{2j}$ for $j=1,...C$,%
\begin{eqnarray*}
&&P(Y=c|Q,S,A,H,W_{0},...,W_{C})= \\
&&\frac{\exp (\Delta \beta _{2c}+\Delta \beta _{\tau c}S+\Delta \beta
_{qc}Q+\Delta \beta _{dc}D+\Delta \beta _{ac}^{\prime }A+\Delta \beta
_{hc}^{\prime }H-\beta _{w0}^{\prime }W_{0}+\beta _{wc}^{\prime }W_{c})}{%
1+\sum_{j=1}^{C}\exp (\Delta \beta _{2j}+\Delta \beta _{\tau j}S+\Delta
\beta _{qj}Q+\Delta \beta _{dj}D+\Delta \beta _{aj}^{\prime }A+\Delta \beta
_{hj}^{\prime }H-\beta _{w0}^{\prime }W_{0}+\beta _{wj}^{\prime }W_{j})}.
\end{eqnarray*}%
Noting $\Delta \beta _{20}=\Delta \beta _{30}=0$, the numerator becomes one
for the base class $c=0$.

\qquad Because $D$ alters the choice probability for class $c$ by $\beta
_{dc}$, the \textquotedblleft net increase\textquotedblright\ in the
propensity to choose class $c$ relative to the base class $0$ is $\Delta
\beta _{dc}\equiv \beta _{dc}-\beta _{d0}$, not $\beta _{dc}$. Estimate $%
\Delta \beta _{d1},...,\Delta \beta _{dC}$ with cross-section multinomial
logit using the last display. Then, $\exp (\Delta \beta _{dc})-1$ is the
class-$c$ proportional odds effect relative to the class $0$, and the class-$%
c$ proportional effect as well when $Y=c$ is a rare event in the sense of
(3.3).\bigskip 

\textit{Simple Simulation Study for Multinomial Response\bigskip }

\qquad Our simulation study using the above $P(Y=c|Q,S,A,H,W_{0},...,W_{C})$
with $C=2$ has the following design (the error terms generated as in (A.4)
and $H_{it}$ excluded):%
\begin{eqnarray*}
A &\sim &Uniform(-1,1),\text{ \ }W_{c2},W_{c3}\text{ for }c=0,1,2\text{ are
iid }N(0,1),\ \ P(Q=1)=0.5\text{,} \\
P(S &=&1)=0.5,\ \ \ \beta _{20}=\beta _{30}=\beta _{q0}=\beta _{d0}=\beta
_{a0}=\beta _{w0}=0\text{ \ (for class 0),} \\
\beta _{21} &=&\beta _{22}=-4,\ \ \beta _{31}=\beta _{32}=-5,\ \ \beta
_{q1}=\beta _{q2}=-0.5\text{ \ (for classes 1,2),} \\
\beta _{d1} &=&\beta _{d2}=0.5,\ \ \beta _{a1}=\beta _{a2}=0.5,\ \ \beta
_{w1}=\beta _{w2}=0.5\text{ \ (for classes 1,2).}
\end{eqnarray*}%
That is, the class-0 parameters are all zero, and the parameters of classes
1 and 2 are the same. Due to $\beta _{20}=\beta _{30}=0$ but $\beta
_{21}=\beta _{22}=-4$ and $\beta _{31}=\beta _{32}=-5$ (much smaller
intercepts for classes 1 and 2 relative to class 1), the events $Y=1,2$ are
rare.

\begin{center}
\begin{tabular}{ccccccc}
\hline\hline
\multicolumn{7}{c}{Table A5. Multinomial $Y$: 3 Classes,\ $N=10,000$, $5,000$
Repetitions} \\ 
& \multicolumn{3}{c}{Class $c=1$} & \multicolumn{3}{c}{Class $c=2$} \\ 
& True, \TEXTsymbol{\vert}Bias\TEXTsymbol{\vert} & SD, RMSE & AvgSE & True, 
\TEXTsymbol{\vert}Bias\TEXTsymbol{\vert} & SD, RMSE & AvgSE \\ \hline
$\Delta \beta _{2c}$ & -4.0,\ 0.019 & 0.15,\ 0.023 & 0.15 & -4.0, 0.030 & 
0.15, 0.024 & 0.15 \\ 
$\Delta \beta _{\tau c}$ & -1.0,\ 0.017 & 0.28, 0.078 & 0.28 & -1.0, 0.011 & 
0.28, 0.077 & 0.28 \\ 
$\Delta \beta _{qc}$ & -0.5,\ 0.004 & 0.23,\ 0.055 & 0.23 & -0.5, 0.006 & 
0.23, 0.054 & 0.23 \\ 
$\Delta \beta _{dc}$ & 0.5,\ 0.004 & 0.41,\ 0.166 & 0.41 & 0.5, 0.007 & 
0.41, 0.168 & 0.41 \\ 
$\Delta \beta _{ac}$ & 0.5,\ 0.003 & 0.16,\ 0.026 & 0.17 & 0.5, 0.005 & 
0.17, 0.027 & 0.17 \\ 
$\beta _{w0}$ & 0.0, 0.001 & 0.07, 0.004 & 0.07 &  &  &  \\ 
$\beta _{wc}$ & 0.5, 0.000 & 0.09, 0.008 & 0.10 & 0.5, 0.001 & 0.09, 0.008 & 
0.10 \\ \hline
\multicolumn{7}{c}{AvgSE is the average of the standard error estimates} \\ 
\hline\hline
\end{tabular}
\end{center}

\qquad Table A5 presents the simulation results, where each entry consists
of true values (True), \TEXTsymbol{\vert}Bias\TEXTsymbol{\vert}, SD, RMSE,
and the average of the standard error estimates (AvgSE). Overall, biases are
very small, and AvgSE's are almost the same as the SD's. With $N=10,000$,
the multinomial logit with RCS works well even for rare events $Y=1,2$%
.\bigskip 

\begin{center}
{\LARGE REFERENCES}
\end{center}

\qquad Ai, C. and E.C. Norton, 2003, Interaction terms in logit and probit
models, Economics Letters 80, 123-129.

\qquad Angrist, J.D. and A.B. Krueger, 1999, Empirical strategies in labor
economics, in Handbook of Labor Economics 3A, edited by O. Ashenfelter and
D. Card, North-Holland.

\qquad Angrist, J.D. and J.S. Pischke, 2009, Mostly harmless econometrics,
Princeton University Press.

\qquad Athey, S. and G.W. Imbens, 2006, Identification and inference in
nonlinear difference-in-differences models, Econometrica 74, 431-497.

\qquad Barbaresco, S., C.J. Courtemanche and Y. Qi, 2015, Impacts of the
Affordable Care Act dependent coverage provision on health-related outcomes
of young adults, Journal of health economics 40, 54-68.

\qquad Cataife, G. and M.B. Pagano, 2017, Difference in difference: simple
tool, accurate results, causal effects, Transfusion 57, 1113-1114.

\qquad Ciani, E. and P. Fisher, 2019, Dif-in-dif Estimators of
multiplicative treatment effects, Journal of Econometric Methods, 20160011.

\qquad Dukes, O. and S. Vansteelandt, 2018, A note on G-estimation of causal
risk ratios, American Journal of Epidemiology 187, 1079-1084.

\qquad Jena, A.B., D.P. Goldman and S.A. Seabury, 2015, Incidence of
sexually transmitted infections after human papillomavirus vaccination among
adolescent females, JAMA Internal Medicine 175, 617-623.

\qquad Kahn-Lang A. and K. Lang, 2020, The promise and pitfalls of
differences-in-differences: reflections on \textit{16 and pregnant} and
other applications, Journal of Business and Economic Statistics 38, 613-620.

\qquad Kim, Y.S. and M.J. Lee, 2017, Ordinal response generalized
difference-in-differences with varying categories: the health effect of a
disability program in Korea, Health Economics 26, 1121-1131.

\qquad Lechner, M., 2011, The estimation of causal effects by
difference-in-difference methods, Foundations and Trends in Econometrics 4,
165-224.

\qquad Lee, M.J., 2005, Micro-econometrics for policy, program, and
treatment effects, Oxford University Press.

\qquad Lee, M.J., 2010, Micro-econometrics: methods of moments and limited
dependent variables, Springer.

\qquad Lee, M.J., 2015, Panel conditional and multinomial logit estimators,
in The Oxford Handbook of Panel Data, 202-232, edited by B. Baltagi, Oxford
University Press

\qquad Lee, M.J., 2016a, Matching, regression discontinuity, difference in
differences, and beyond, Oxford University Press.

\qquad Lee, M.J., 2016b, Generalized difference in differences with panel
data and least squares estimator, Sociological Methods \& Research 45,
134-157.

\qquad Lee, M.J., 2018, Simple least squares estimator for treatment effects
using propensity score residuals, Biometrika 105, 149-164.

\qquad Lee, M.J., 2021, Instrument residual estimator for any response
variable with endogenous binary treatment, Journal of the Royal Statistical
Society (Series B) 83, 612-635.

\qquad Lee, M.J. and Y.S. Kim, 2014, Difference in differences for stayers\
with a time-varying qualification:\ health expenditure elasticity of the
elderly, Health Economics 23, 1134-1145.

\qquad Lee, M.J. and S. Kobayashi, 2001, Proportional treatment effects for
count response panel data: effects of binary exercise on health care demand,
Health Economics 10, 411-428

\qquad Lee, M.J. and Y. Sawada, 2020, Review on difference in differences,
Korean Economic Review 36, 135-173.

\qquad McGrath S.P., I.M. Perreard, M.D. Garland, K.A. Converse and T.A.
Mackenzie, 2019, Improving patient safety and clinician workflow in the
general care setting with enhanced surveillance monitoring, IEEE Journal of
Biomedical and Health Informatics 23, 857-866.

\qquad Morgan, S.L. and C. Winship, 2014, Counterfactuals and causal
inference, 2nd ed., Cambridge University Press.

\qquad Nelder, J.A. and R.W.M. Wedderburn, 1972, Generalized linear models,
Journal of the Royal Statistical Society (Series A) 135, 370-384.

\qquad Papke, L.E. and J.M. Wooldridge, 1996, Econometric methods for
fractional response variables with an application to 401 (k) plan
participation rates, Journal of Applied Econometrics 11, 619-632.

\qquad Puhani, P.A., 2012, The treatment effect, the cross difference, and
the interaction term in nonlinear \textquotedblleft difference in
differences\textquotedblright\ models, Economics Letters 115, 85-87.

\qquad Santos Silva, J.M.C. and S. Tenreyro, 2006, The log of gravity,
Review of Economics and Statistics 88, 641-658.

\qquad Santos Silva, J.M.C. and S. Tenreyro, 2011, Further simulation
evidence on the performance of the Poisson pseudo-maximum likelihood
estimator, Economics Letters 112, 220-222.

\qquad Shadish, W.R., T.D. Cook and D.T. Campbell, 2002, Experimental and
quasi-experimental designs for generalized causal inference, Houghton
Mifflin Company.

\qquad Yadlowsky, S., F. Pellegrini, F. Lionetto, S. Braune and L. Tian,
2021, Estimation and validation of ratio-based conditional average treatment
effects using observational data, Journal of the American Statistical
Association 116, 335-352.

\end{document}